\documentclass[conference]{IEEEtran}
\IEEEoverridecommandlockouts
\usepackage{cite}
\usepackage{amsmath,amssymb,amsfonts}
\usepackage{algorithmic}
\usepackage{graphicx}
\usepackage{textcomp}
\usepackage{xcolor}
\def\BibTeX{{\rm B\kern-.05em{\sc i\kern-.025em b}\kern-.08em
    T\kern-.1667em\lower.7ex\hbox{E}\kern-.125emX}}

\usepackage{pifont}
\newcommand*\circled[1]{\tikz[baseline=(C.base)]{
    \node[shape=circle,draw,inner sep=0.1pt] (C) {#1};}}

\usepackage{soul}
\usepackage{array}
\usepackage{enumitem}
\usepackage{multirow}
\usepackage{setspace}
\usepackage{comment}
\usepackage[]{siunitx}
\usepackage{wrapfig,lipsum}
\usepackage{xspace}
\usepackage{tikz}
\usepackage{subcaption}

\usepackage{hyperref}

\newcommand{\pimmalloc}[0]{PIM-malloc\xspace}
\newcommand{\pimmallocSW}[0]{PIM-malloc-SW\xspace}
\newcommand{\pimmallocHW}[0]{PIM-malloc-HW/SW\xspace}

\newcommand{\fig}[1]{Figure~\ref{#1}}

\newcommand{\sect}[1]{Section~\ref{#1}}
\newcommand{\tab}[1]{Table~\ref{#1}}

\begin{document}

\title{\fontsize{22}{22}\selectfont\pimmalloc: A Fast and Scalable Dynamic Memory Allocator for Processing-In-Memory (PIM) Architectures\thanks{\IEEEauthorrefmark{1} Co-first authors who contributed equally to this research.}\thanks{\IEEEauthorrefmark{2} Work done while at KAIST.}
}

\author{\IEEEauthorblockN{Dongjae Lee\IEEEauthorrefmark{1}}
\IEEEauthorblockA{
KAIST\\
dongjae.lee@kaist.ac.kr}
\and
\IEEEauthorblockN{Bongjoon Hyun\IEEEauthorrefmark{1}\IEEEauthorrefmark{2}}
\IEEEauthorblockA{
Samsung Electronics\\
bongjoon.hyun@gmail.com}
\and
\IEEEauthorblockN{Youngjin Kwon}
\IEEEauthorblockA{
KAIST\\
yjkwon@kaist.ac.kr}
\and
\IEEEauthorblockN{Minsoo Rhu}
\IEEEauthorblockA{
KAIST\\
mrhu@kaist.ac.kr}
}

\maketitle

\begin{abstract}
The ability to dynamically allocate memory is fundamental in modern programming languages. However, this feature is not adequately supported in current general-purpose PIM devices. To identify key design principles that PIM must consider, we conduct a design space exploration of PIM memory allocators, examining various strategies for metadata placement and management of the allocator. Based on this exploration, we introduce \pimmalloc, a fast and scalable memory allocator for general-purpose PIM that operates on real PIM hardware, achieving a $66\times$ improvement in memory allocation performance. This design is further enhanced with a lightweight, per-PIM core hardware cache, specifically designed for dynamic memory allocation, achieving an additional $31\%$ performance improvement. Finally, we demonstrate the applicability of \pimmalloc by developing several representative PIM workloads, demonstrating its effectiveness in enhancing programmability.
\end{abstract}


\section{Introduction}
\label{sect:introduction}
Recent years have seen renewed interest in Processing-In-Memory (PIM) architectures~\cite{diva, tom, a_modern_primer_on_processing_in_memory, pim_enabled_instructions, conda, lazypim, tensor_dimm, recnmp, tetris, neurocube, mcn, impica, dimmining, polynesia, pimcloud, accelerating_neural_network_inference_with_processing_in_dram, napel, spacea, dracc, gearbox, natsa, aespa, drisa, cairo, tensor_casting, a_logic_in_memory_computer, near_data_acceleration_with_concurrent_host_access, affinity_alloc, fafnir, charon, stream_based_ndp_extended, gao2016hrl, tian2024ndpbridge, tian2023abndp, nai2018coolpim, boroumand2018google, duplex, psyncpim, fujiki2018memory, pattnaik2016scheduling}, with several commercial PIM systems now available. Memory vendors offer \emph{domain-specific} PIM solutions targeting application-specific compute primitives~\cite{hc_aim_new, hc_samsung_new, cxl_pnm, hbm_pim_isca, hbm_pim_isscc, newton, an_fpga_based_rnn_t_inference_accelerator_with_pim_hbm, mcdram}, whereas UPMEM’s solution~\cite{upmem_hotchips} provides a more flexible, \emph{general-purpose} programming language that enables developers to write \emph{any} parallel program for PIM execution. This high degree of programmability has spurred extensive research, broadening the reach of general-purpose PIM across diverse application domains~\cite{prim, prim_2, prim_3, a_case_study_of_processing_in_memory_in_off_the_shelf_systems, bulk_jpeg_decoding_on_in_memory_processors, upmem_sigmod, jk_sigmetric, sparsep, sparsep_arxiv, upmem_ispass_ml, upmem_isvlsi_ml, upmem_arxiv_ml, gogineni2024swiftrl, rhyner2024analysis, rhyner2024pim, swiftrl, homomorphic_upmem, spid, atim}. Given this landscape, this paper focuses on system software support, specifically dynamic memory allocation for commodity general-purpose PIM, a capability that remains insufficiently supported in current PIM devices. We observe that the following two properties of PIM pose unique challenges in designing a high-performance PIM memory allocator:

\textbf{(Challenge \#1: Explosion of memory address spaces to manage)}
Contemporary PIM systems are based on a \emph{bank-level} PIM architecture, where each memory bank is paired with a dedicated PIM core~\cite{upmem_hotchips}. While this design effectively enhances overall memory bandwidth, each per-bank PIM core is limited to accessing only the data stored within its respective local DRAM bank. As a result, the PIM system must manage thousands of \emph{distinct} memory address spaces, one for each PIM core. For instance, an UPMEM-PIM system with 2,560 PIM cores requires the management of 2,560 separate memory address spaces, each necessitating its own heap space and the corresponding metadata to track allocated data, e.g., allocated memory block size and free lists. Compared to CPUs and GPUs, which have only a \emph{single} memory address space to manage, such ``explosion'' of address spaces in PIM systems poses unique challenges, as decisions must be made regarding \emph{where} to store the thousands of sets of metadata for memory allocation (e.g., all in the centralized CPU memory vs. partitioned across each PIM core’s local memory).

\textbf{(Challenge \#2: Wimpy PIM cores that execute the memory allocation algorithm)} A typical PIM system contains a \emph{heterogeneous} mix of CPU cores and PIM cores, each with its own local memory address space. CPU cores are designed using the latest performance-optimized logic process technology, allowing them to operate in the GHz frequency range with sophisticated microarchitectures. In contrast, current-generation PIM cores are significantly less powerful because they are fabricated using a DRAM processing technology node, operating in the several hundreds of MHz range~\cite{upmem_hotchips, hbm_pim_isca, hbm_pim_isscc}. To compensate for the reduced performance of each PIM core, PIM systems integrate thousands of PIM cores to maximize throughput. This disparity in performance characteristics raises a critical question regarding \emph{which processor} (e.g., CPU cores vs. PIM cores) is optimal for executing the sophisticated memory allocation algorithm, as this choice impacts the speed of memory allocation.

Considering these critical design challenges, our first important contribution is a thorough examination and characterization of the design space of dynamic memory allocators for PIM. Based on this exploration, we design, implement, and evaluate a fast and scalable dynamic memory allocator for PIM architectures, which we henceforth refer to as \emph{\pimmalloc}. Our \pimmalloc design comes in two different versions. The \emph{software-only} \pimmalloc is designed to be executable on real-world, programmable PIM devices like UPMEM-PIM. Compared to a naively designed PIM memory allocator that does not consider the unique properties of PIM, our software-only \pimmalloc performs $66\times$ faster with high scalability. We also present our \emph{hardware/software co-designed} \pimmalloc, which provides further performance benefits on top of our software-only \pimmalloc. By allocating a small, per-PIM core hardware cache, our proposed \pimmalloc function is able to achieve an additional $31\%$ performance improvement compared to our software-only \pimmalloc design. Overall, we believe that \pimmalloc will serve as a foundation for future research, enabling PIM programmers to explore a broader spectrum of applications and unlock the full potential of general-purpose PIM\footnote{Open-sourced at \url{https://github.com/VIA-Research/PIM-malloc}.}.

\section{Background}
\label{sect:background}

\subsection{UPMEM-PIM: A Programmable PIM System}
\label{sect:background_upmem}

\begin{figure}[t!] \centering
%\vspace{-1.3em}
\includegraphics[width=0.40\textwidth]{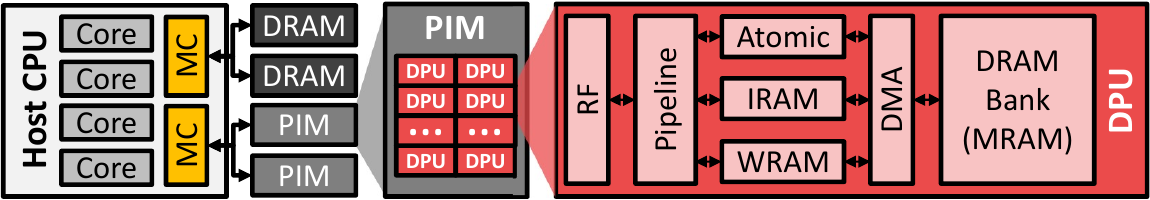} 
%\vspace{-0.5em}
  \caption{UPMEM-PIM hardware system overview.}
\label{fig:upmem_pim_enabled_memory_overview}
\vspace{-1.3em}
\end{figure}

{\bf Hardware architecture.} UPMEM-PIM comprises a host CPU connected to both standard DRAM and PIM modules via the memory bus (\fig{fig:upmem_pim_enabled_memory_overview}). Each PIM module adheres to the DDR4-2400~\cite{ddr4_2400} protocol and DIMM form factor, housing $128$ DRAM Processing Units (DPUs). The DPU itself is an in-order, RISC-based processor clocked at $350$ MHz. These DPUs are tightly coupled with a $64$ KB scratchpad memory (WRAM), $24$ KB of instruction memory (IRAM), and a dedicated, local DRAM bank (MRAM). Leveraging fine-grained multithreading, each DPU supports the concurrent execution of up to $24$ threads. While the $24$ threads can share access to a DPU's local MRAM, WRAM, and IRAM, they are not able to address data located in other DPUs. In other words, a given DPU has its own \emph{local} memory address space (i.e., each DPU's local DRAM bank, its own WRAM and IRAM), rendering the overall PIM system to end up having $N$ \emph{distinct} memory address spaces managed independently across $N$ different DPUs. This is in stark contrast to CPUs or GPUs where multiple CPU/GPU cores share a \emph{single} memory address space (e.g., global memory in CUDA).

{\bf Software system and programming model.}
UPMEM-PIM programming model is based on the single-program multiple-data (SPMD) paradigm, requiring careful data partitioning across DPUs and threads. 
An LLVM~\cite{llvm, upmem_llvm}-based compiler toolchain facilitates the translation of UPMEM programs (written using a C-like programming language provided by UPMEM) into DPU machine code. UPMEM-PIM employs a co-processor model, reminiscent of NVIDIA's CUDA~\cite{cuda}, wherein a host CPU delegates memory-intensive tasks to the DPUs integrated within the memory subsystem. This architecture necessitates the development of two distinct binaries: one for the host CPU and another for the DPUs. The host CPU is in charge of DPU allocation, program binary distribution, data partitioning and transfer, initiation of PIM program execution, and retrieval of results. The UPMEM-PIM programming model is characterized by its \emph{scratchpad-centric} memory architecture. This design requires the programmer to first explicitly transfer the working set from DRAM (MRAM) to the scratchpad (WRAM). Once the data is uploaded to the scratchpad, threads can access it using load/store instructions.

{\bf Dynamic memory allocation in UPMEM-PIM.}
UPMEM-PIM currently does not support dynamic memory allocation in its local DRAM bank (MRAM). To simplify programming, UPMEM-PIM offers a limited form of memory allocation in the 64 KB scratchpad memory (WRAM) using an API called \texttt{buddy\_alloc()}. This function employs the \emph{buddy allocation} algorithm~\cite{buddy_alloc} to enable dynamic allocation and deallocation of blocks with varying sizes. However, limiting memory allocation to such small 64 KB region significantly constrains programmability as well as user productivity.

\begin{figure}[t!] \centering
  \includegraphics[width=0.475\textwidth]{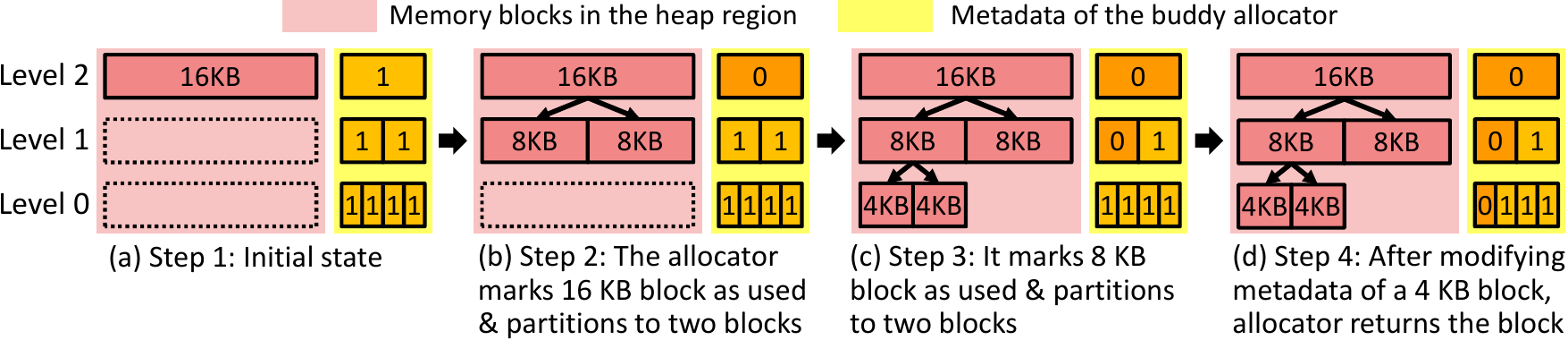}
  \caption{An example workflow of buddy allocator.}
  \label{fig:buddy_allocator}
\vspace{-1.3em}
\end{figure}

\subsection{Buddy Memory Allocation}
\label{sect:buddy_alloc}
The buddy allocation algorithm organizes memory in power-of-2 block sizes, allowing blocks to split or merge with \emph{buddy} blocks when (de)allocating memory. It helps reduce fragmentation by consolidating free blocks of the same size, preserving large contiguous memory areas. The buddy allocator manages page-level allocations in the kernel, providing a foundation for other allocators (e.g., slab allocator~\cite{linux_slab}).

In memory allocators, \emph{metadata} refers to the additional information stored alongside allocated memory blocks to help the allocator manage memory efficiently. This metadata tracks details about each memory block, such as its size and status (allocated or free). In the example workflow of the buddy memory allocator (\fig{fig:buddy_allocator}), a memory request is made for a $4$ KB chunk while the buddy allocator manages a $16$ KB memory pool. Since the $16$ KB block exceeds the requested size, the allocator initially divides it into two equal buddies at level $2$. It then selects one of these buddies (level $1$, Left) and further divides it into smaller blocks. This recursive division process continues until a $4$ KB block is reached, meeting the requested memory size, and the allocator assigns one of the resulting buddies (level $0$, Left). 

The metadata overhead of the buddy allocator grows rapidly with larger heap and finer allocation granularity. For example, UPMEM‑PIM’s \texttt{buddy\_alloc()} oversees only $32$ KB of scratchpad heap and needs less than 512 B of metadata. In contrast, applying a vanilla buddy allocation algorithm to manage a PIM core's 32 MB of per-bank DRAM heap space incurs 512 KB of metadata for each PIM core, amounting to more than 1 GB of metadata across the 2,560 PIM cores.

\section{Motivation}
\label{sect:motivation}

\begin{figure}[t!] \centering
\includegraphics[width=0.475\textwidth]{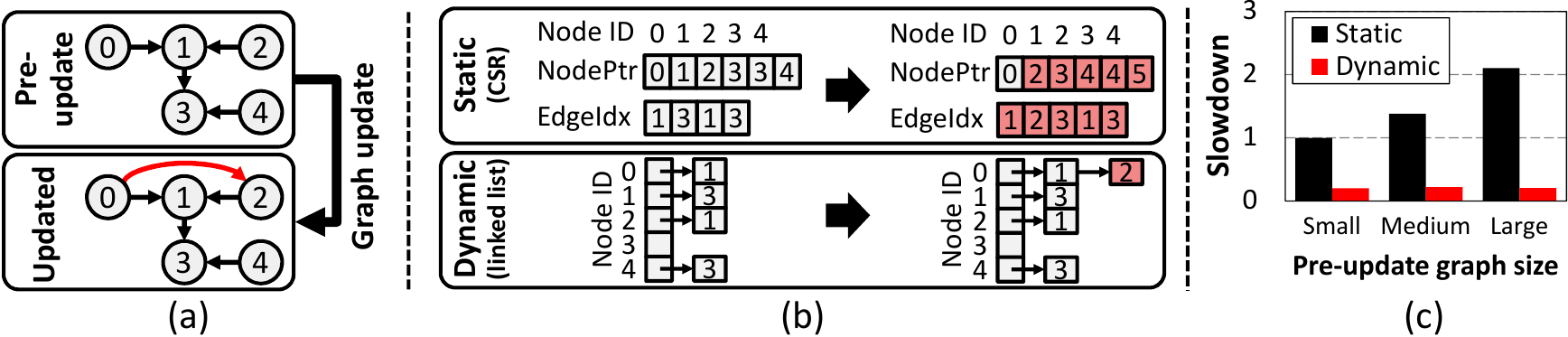} 
  \caption{Illustration of (a) a dynamic graph update operation, (b) a comparison of how static and dynamic data structures are used to manage the graph update shown in (a), and (c) the performance slowdown (normalized to static/small) as the size of the pre-update graph is increased,  from small to large, while the total number of newly added graph edges remains constant. The graph updates using dynamic data structures are implemented using our proposed \pimmallocSW APIs, discussed in detail in \sect{sect:proposed_sw_only}. The loc-gowalla~\cite{cho2011friendship} dataset is used to create the pre-update graph data.  }
  \label{fig:why_malloc}
  \vspace{-1.3em}
\end{figure}

\subsection{The Need for Dynamic Memory Allocators in PIM}
\label{sect:why_malloc_for_pim}
Dynamic memory allocation is a fundamental component of modern programming, allowing flexible and efficient management of memory resources. However, current programming models for general-purpose PIM lack this functionality (i.e., memory allocation is only supported limitedly at the small scratchpad in UPMEM-PIM), which not only complicates PIM programming but can also cause excessive memory consumption. Below, we present two concrete examples that illustrate the inefficiencies that stem from this limitation.

\textbf{(Case study \#1: Dynamic graph update)} Graph analytics, which is typically memory-bound and well-suited for PIM~\cite{graph_p, graph_h, graph_q, tesseract, nai2017graphpim, sisa}, serves as a compelling example. Most real-world graphs are dynamic and change their structure over time, so updating the graph data should not become a performance bottleneck. Existing CPU/GPU-based graph analytics frameworks mitigate such performance overhead by employing dynamic data structures (i.e., those that can change their size at runtime, such as linked lists) for high-performance graph updates~\cite{aspen, terrace, awad2020dynamic, sagabench, faimgraph, basak2021improving}. Unfortunately, current PIM devices do not adequately support dynamic memory allocation and therefore cannot utilize dynamic data structures. Consequently, developers must rely on static data structures, such as compressed sparse row (CSR), which is a fixed-size format that cannot be resized at runtime. This limitation introduces significant drawbacks, which we detail below using \fig{fig:why_malloc}. 

Managing graph updates using static data structures requires cumbersome array operations.  For example, adding a single edge from node 0 to node 2 (red arrow in \fig{fig:why_malloc}(a)) necessitates updating the entire \texttt{NodePtr} array after the first index (top of \fig{fig:why_malloc}(b)). 
Additionally, the \texttt{EdgeIdx} array must be resized and shifted. More critically, manual memory management and pointer manipulation--or, in some cases, rebuilding the entire array--place a significant burden on programmers. In contrast, a dynamic data structure would allow memory to be allocated solely for the new edges and connected via pointers (bottom of \fig{fig:why_malloc}(b)), eliminating labor-intensive updates and improving programmer productivity. \fig{fig:why_malloc}(c) demonstrates how graph update performance changes as the size of the \emph{pre-update graph} (the graph structure \emph{before} the update, shown at the top of \fig{fig:why_malloc}(a)) increases (from \texttt{small} graphs to \texttt{large} graphs), while the total number of newly added graph edges remains fixed. 
When the pre-update graph is stored in a static CSR format, each edge insertion requires reorganizing the array, resulting in a worst-case time complexity that is proportional to the total size of the graph. In contrast, a dynamic data structure only requires updating the pointer variable that references the newly added data. As a result, the size of the pre-update graph has a negligible impact on graph update performance.

\begin{figure}[t!] \centering
%\vspace{-1.3em}
\includegraphics[width=0.475\textwidth]{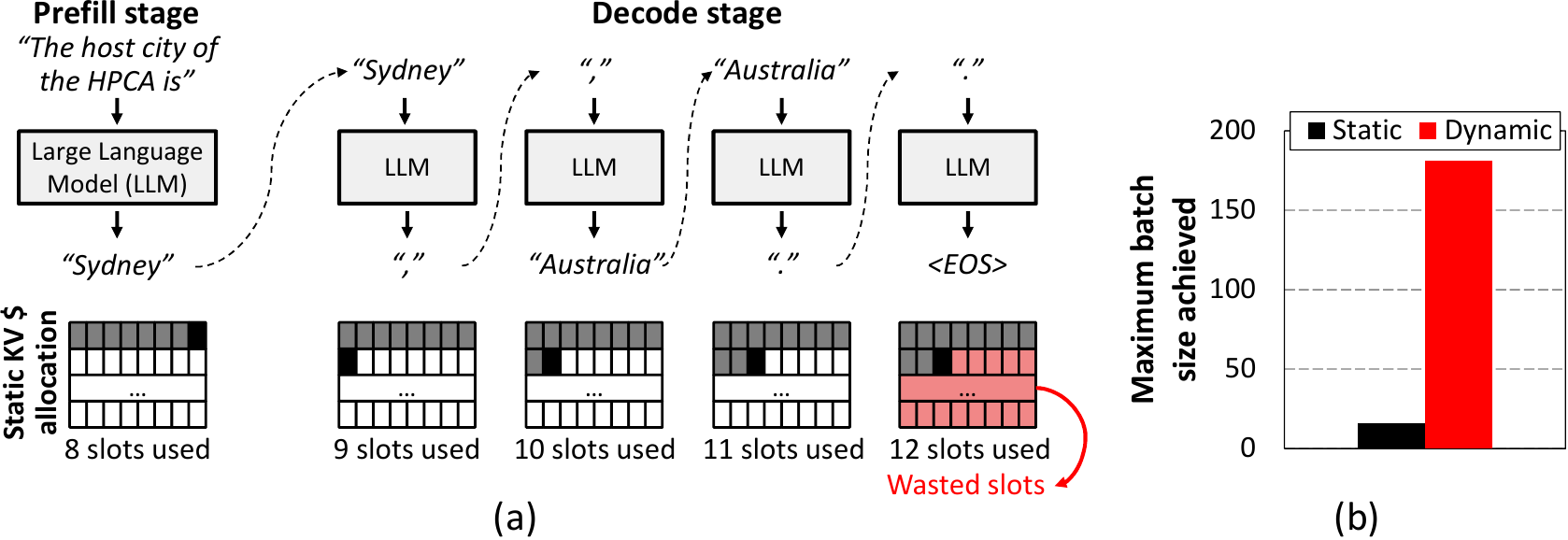} 
  \caption{(a) Illustration of the LLM inference workflow assuming the KV cache is managed with static allocation. (b) Maximum batch size under static and dynamic memory allocation. Measurements are taken on a PIM system equipped with 512 UPMEM‑PIM cores, using the ShareGPT dataset~\cite{sharegpt} and the Llama2‑7B model configuration~\cite{llama2}. \sect{sect:methodology} details our implementation of attention layers.
  }
  \label{fig:why_malloc2}
  \vspace{-1.5em}
\end{figure}

\textbf{(Case study \#2: Attention layer in LLM inference)} 
Large language models (LLMs) primarily consist of fully connected (FC) and attention layers, and their inference proceeds in two stages. (1) \emph{Prefill}: the entire input prompt is processed in one pass, producing key–value (KV) vectors for all tokens and storing them in the KV cache. (2) \emph{Decode}: it then generates output tokens auto‑regressively, appending each new KV vector to the cache (\fig{fig:why_malloc2}(a)). The decode stage dominates inference latency because its sequential dataflow is memory‑bandwidth‑bound, whereas prefill is highly parallel and compute‑intensive~\cite{attacc}. As such, state-of-the-art LLM serving frameworks~\cite{orca,vllm} batch requests to enhance the arithmetic intensity of the decode stage, which helps speed up FC layers but still leaves attention layers memory‑bandwidth‑limited as each request must read and write its own KV cache.

To address the memory-bandwidth limits of attention layers, prior work~\cite{attacc, attacc_cal, neupim, pim_all_you_need, papi, specpim, paise} has proposed offloading attention layers to PIM while executing the remaining layers on the host processor (xPU). In such xPU+PIM systems, achieving high throughput via batching depends on minimizing the allocation size of the KV cache in PIM because KV cache storage grows linearly with both sequence length and batch size; large KV caches limit the maximum batch size that can be formed.
Because the output sequence length is non‑deterministic and known only at runtime, LLM serving frameworks adopt one of two strategies for managing the KV cache. First, the framework statically reserves a large contiguous memory space and then directly manages that region through a \emph{customized memory allocator} (e.g., the KV block manager in vLLM~\cite{vllm}). Second, it can request additional memory on demand by invoking \emph{native memory allocation functions} (e.g., CUDA Virtual Memory Management~\cite{vmm} in vAttention~\cite{vattention}). Because the current PIM system software lacks adequate dynamic memory allocation support, PIM programmers must either undertake the cumbersome task of implementing a customized memory allocator from scratch or \emph{statically} reserve a buffer large enough to accommodate the maximum possible number of KV caches that may be required, leading to excessive memory waste~\cite{paise}. \fig{fig:why_malloc2}(b) compares the maximum batch size achievable when the KV cache is managed with different memory allocation techniques. Static allocation, as implemented in PAISE~\cite{paise}, is constrained to a smaller maximum batch size due to poor memory utilization from fragmentation. In contrast, dynamic allocation allocates KV‑cache space economically at runtime, minimizes fragmentation, and enables larger batches.

\subsection{Design Space of Dynamic Memory Allocators in PIM}
\label{sect:dse_buddy_alloc}
In this section, we seek to provide answers to the following \textbf{two fundamental questions} that highlight the unique design challenges of dynamic memory allocators for PIM.

\textbf{Q1: Where should the metadata be stored?} Modern PIM systems use a \emph{bank-level} PIM architecture, where each DRAM bank is paired with a dedicated PIM core. This setup significantly improves aggregate memory bandwidth available to the PIM system but restricts each PIM core to accessing data only within its local DRAM bank. Consequently, the PIM system must manage thousands of \emph{separate} memory address spaces across all PIM cores, meaning an independent set of metadata must be maintained for each address space. For instance, UPMEM-PIM with 2,560 PIM cores requires managing 2,560 independent address spaces, each with its own heap space and metadata for tracking allocated data. Unlike traditional CPUs and GPUs which operate within a single address space, PIM systems present unique design challenges due to the need to manage numerous address spaces. This raises a critical question about \emph{where to store the system-wide metadata for dynamic memory allocators}: either in the host CPU in a \emph{centralized} manner, or \emph{distributed} across the PIM cores, where each PIM core independently maintains the heap space and metadata locally within its respective DRAM bank.

\begin{table}[t!] \centering
  \caption{Design space of PIM memory allocators.}
\footnotesize
\resizebox{0.43\textwidth}{!}{%
  \begin{tabular}{c c c}
    \hline
        \textbf{Design}            & \textbf{Metadata}           &\textbf{Processor executing} \\
        \textbf{strategy}          & \textbf{storage location}   &\textbf{ memory alloc. algorithm} \\
    \hline
    \hline
   	Host-Metadata/      	& (Centralized),    & Tens of \\
		Host-Executed 		    & host CPU memory   &  brawny CPU cores \\
    \hline                              
        Host-Metadata/	    	& (Centralized)     & Hundreds of \\
        PIM-Executed            & host CPU memory   & wimpy PIM cores  \\
	\hline
   	PIM-Metadata/	        & (Distributed)	   & Tens of \\
        Host-Executed           & PIM memory       & brawny CPU cores  \\
	\hline
   	PIM-Metadata/	        & (Distributed)	   & Hundreds of \\
        PIM-Executed            & PIM memory       & wimpy PIM cores  \\
    \hline
  \end{tabular}%
  }
\vspace{-1.3em}
  \label{tab:design_space_exploration}
\end{table}

\textbf{Q2: Which processor architecture should retain ownership of executing the memory allocation algorithm?} PIM system comprises of two main compute units: host processor and PIM cores, which differ significantly in architecture and performance characteristics. In UPMEM-PIM, for instance, the CPU serves as the host processor which contains multiple high-performance cores operating at several GHz of operating frequency, with advanced microarchitectures like branch predictors and prefetchers that enable high-performance processing. In contrast, PIM cores, often numbering in the hundreds to thousands per system, are optimized for system-wide parallel computations but operate individually at much lower clock frequencies due to limitations of DRAM process technology. For example, the PIM cores in UPMEM-PIM use an in-order pipeline design and lack microarchitectural features that exploit instruction-level parallelism for high performance. This large discrepancy in processor architecture characteristics creates noticeable trade-offs and raises interesting questions about \emph{which processor architecture is most optimal for executing the dynamic memory allocation algorithm}.

\begin{figure}[t!] \centering
\includegraphics[width=0.475\textwidth]{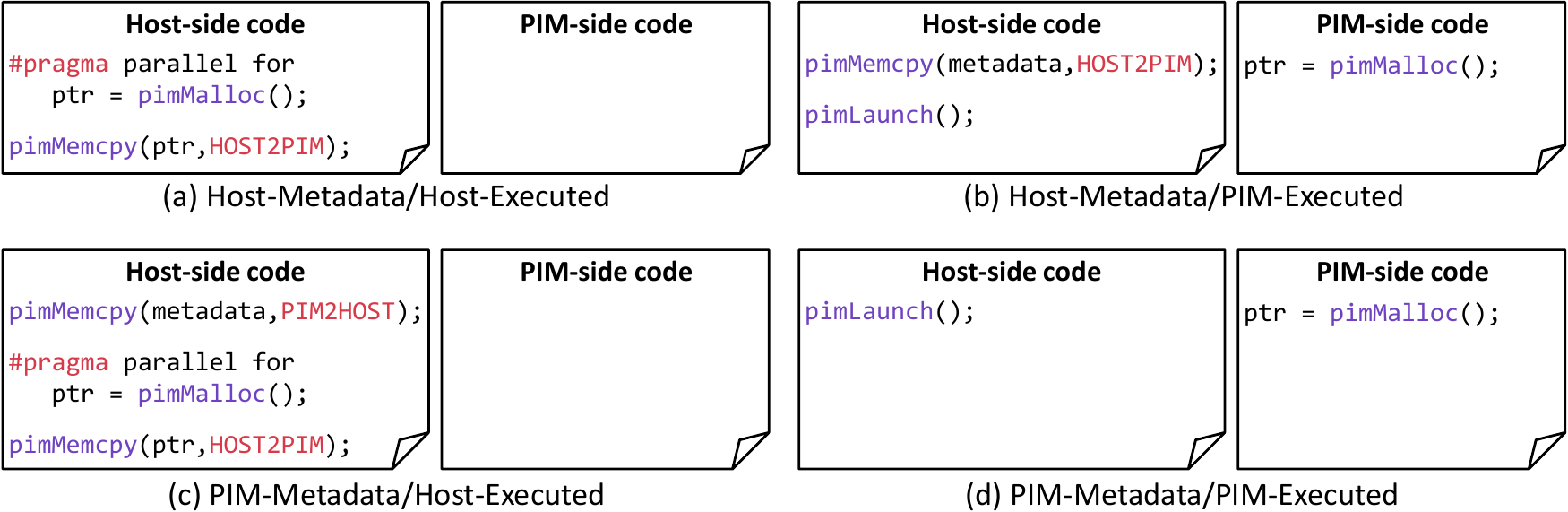} 
  \caption{Pseudo-code of how the design strategies in \tab{tab:design_space_exploration} can be utilized to implement our straw-man \texttt{buddy\_alloc\_PIM\_DRAM} design. Pseudo-code features three primary functions: \texttt{pimMalloc()} allocates memory within the heap space in PIM's DRAM banks; \texttt{pimMemcpy()} (host-side code only) handles DRAM$\leftrightarrow$PIM data transfer; \texttt{pimLaunch()} (host-side code only) instructs PIM cores to execute the buddy allocation algorithm on the PIM side.}
  \label{fig:dse_pseudocode}
  \vspace{-1.3em}
\end{figure}

Based on the aforementioned two key design parameters, we classify the design space of PIM memory allocators into four approaches summarized in \tab{tab:design_space_exploration}. For our design space exploration, we extend UPMEM-PIM’s existing scratchpad-based buddy allocator (i.e., \texttt{buddy\_alloc()} in \sect{sect:buddy_alloc}) to support (de)allocation within each PIM core's local DRAM bank, which we henceforth refer to as the \texttt{buddy\_alloc\_PIM\_DRAM} design. In \texttt{buddy\_alloc\_PIM\_DRAM}, each PIM core is given a 32 MB heap space managed with a 20-level tree metadata ($log_2(32 \ MB \ / \ 32 \ bytes) = 20$), allowing a minimum allocation size of 32 bytes. In the case of \texttt{buddy\_alloc\_PIM\_DRAM}'s design point executed on the PIM core (``PIM-Executed''), recall from \sect{sect:background_upmem} that UPMEM-PIM employs a scratchpad-centric programming model, which requires the working set to be uploaded from DRAM (MRAM) to the scratchpad (WRAM) before use. Because the size of \texttt{buddy\_alloc\_PIM\_DRAM}'s metadata exceeds the scratchpad capacity, our implementation of \texttt{buddy\_alloc\_PIM\_DRAM} (``PIM-Executed'') maintains a \emph{software-managed metadata buffer} in the scratchpad. This buffer is designed to exploit locality by caching recently accessed metadata and its neighboring entries. During the buddy allocator's tree traversal, a miss in this software-managed buffer triggers a metadata fetch operation, transferring a contiguous block of metadata from DRAM to its buffer. We utilize \texttt{buddy\_alloc\_PIM\_DRAM} as a \emph{straw-man design} to underscore the research challenges addressed by our proposal.

\fig{fig:dse_pseudocode} provides pseudo-code illustrating how the four design strategies in \tab{tab:design_space_exploration} can be utilized to implement the straw-man \texttt{buddy\_alloc\_PIM\_DRAM}. For cases where the buddy allocator is executed by the PIM core  (``PIM-Executed''), we implement the buddy allocation algorithm using UPMEM-PIM’s C-based programming language, allowing each PIM core to independently handle PIM memory allocation requests (\texttt{pimMalloc()} in \fig{fig:dse_pseudocode}(b,d)). In such scenarios, the host CPU must explicitly request the target PIM cores to execute the \texttt{pimMalloc()} functions, so a separate host-side \texttt{pimLaunch()} function is used for ``PIM-Executed'' approaches (\fig{fig:dse_pseudocode}(b,d)). As for ``Host-Executed'' scenarios, the CPU cores execute the buddy allocation algorithm, updating each PIM core's metadata to reflect the heap status after servicing the PIM memory allocation request. To maximize performance, we use the \texttt{pthreads} library~\cite{pthread} to parallelize the execution of the buddy allocator's metadata updates (\texttt{parallel for pimMalloc()} in \fig{fig:dse_pseudocode}(a,c)). 

\begin{figure}[t!] \centering
%\vspace{-1.3em}
  \includegraphics[width=0.475\textwidth]{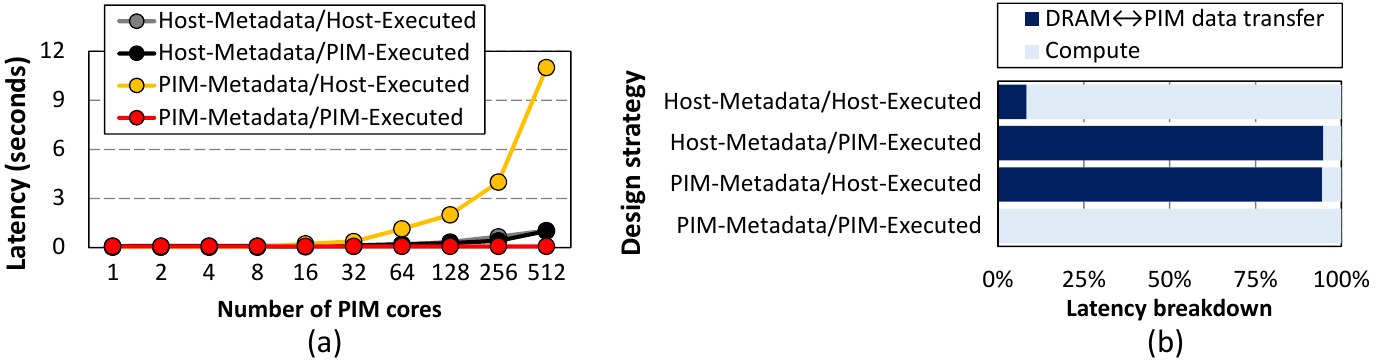}
  %\vspace{-1.3em}
  \caption{(a) Average PIM memory allocation latency when employing the four design strategies discussed in \tab{tab:design_space_exploration}. This experiment assumes a scenario where 1 to 512 PIM cores request an identical 128 memory allocations with the size of 32 bytes concurrently. (b) Latency breakdown with 512 PIM cores data point in (a).}
  \label{fig:design_space_exploration}
  \vspace{-1.3em}
\end{figure}

Note that in the four design approaches shown in \fig{fig:dse_pseudocode}, there are two cases where the host CPU and PIM cores must exchange important information to support \texttt{pimMalloc()}'s execution. First, in ``Host-Executed'' approaches, the pointer values (\texttt{ptr}) returned by a \texttt{pimMalloc()} call must be copied over to the corresponding PIM cores (\texttt{HOST2PIM}). This ensures that the PIM cores can utilize these pointers during their program execution (\fig{fig:dse_pseudocode}(a,c)). 
Second, when metadata storage and buddy allocator execution occur in different locations  (``Host-Metadata/PIM-Executed" and ``PIM-Metadata/Host-Executed"), an explicit copy of the \texttt{metadata} must be exchanged between the host and PIM cores (\texttt{HOST2PIM} in \fig{fig:dse_pseudocode}(b) and \texttt{PIM2HOST} in (c)). These host$\leftrightarrow$PIM data transfers are conducted using \texttt{pimMemcpy()}, implemented with UPMEM-PIM's runtime data transfer API (e.g., \texttt{dpu\_push\_xfer()}).

In \fig{fig:design_space_exploration}(a), we show how memory allocation latency changes as the number of PIM cores requesting allocations increases. Notably, with the exception of ``PIM-Metadata/PIM-Executed'' (red), allocation latency increases for all other design points as the number of PIM cores grows. Both ``Host-Metadata/PIM-Executed'' (black) and ``PIM-Metadata/Host-Executed'' (yellow) designs require metadata to be transferred to the processors executing the buddy allocation algorithm, involving explicit data transfers between standard DRAM and PIM memory. Since metadata size scales with the number of PIM cores, the data transfer overhead becomes substantial, resulting in increased system-wide memory allocation latency (\fig{fig:design_space_exploration}(b)). Although ``Host-Metadata/Host-Executed'' (gray) avoids excessive data movement between DRAM and PIM, it demonstrates limited scalability due to the CPU's constrained parallelism. In contrast, despite relying on wimpy PIM cores, ``PIM-Metadata/PIM-Executed'' (red) provides a highly scalable solution. This is because ``PIM-Metadata/PIM-Executed'' eliminates data transfer overhead between DRAM and PIM while allowing each PIM core to handle its allocation requests locally and in parallel. Overall, we conclude that the ``PIM-Metadata/PIM-Executed'' approach is the optimal baseline design point for implementing a high-performance PIM memory allocator. For the remainder of this paper, both the straw-man \texttt{buddy\_alloc\_PIM\_DRAM} and our proposed \pimmalloc design are based on this design principle.

\subsection{Remaining Challenges with the Straw-man PIM Allocator}
\label{sect:challenge_strawman}

\begin{figure}[t!] \centering
  \includegraphics[width=0.46\textwidth]{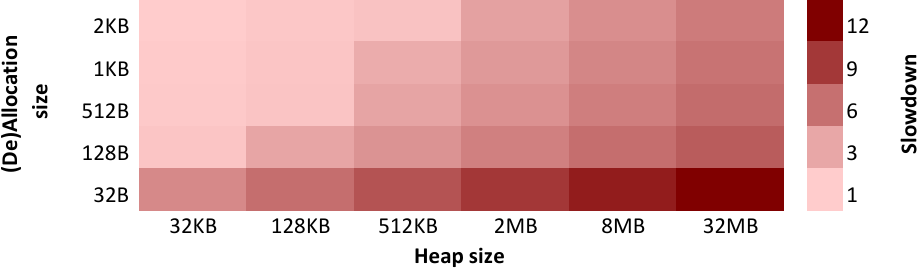}
  \caption{Performance slowdown in the straw-man PIM buddy allocator when the heap size and memory (de)allocation size is varied. Performance is normalized to the case when heap size is 32 KB and allocation size is 2 KB. Performance is measured using a single-threaded UPMEM-PIM program that performs consecutive memory (de)allocation using \texttt{buddy\_alloc\_PIM\_DRAM} which we utilize to calculate average latency.}
  \label{fig:fast_and_slow_paths}
  \vspace{-1.3em}
\end{figure}

While our design space exploration in \sect{sect:dse_buddy_alloc} identified the ``PIM-Metadata/PIM-Executed'' approach as a strong foundation for the straw-man \texttt{buddy\_alloc\_PIM\_DRAM} design\footnote{
We interchangeably refer to the ``PIM-Metadata/PIM-Executed'' straw-man \texttt{buddy\_alloc\_PIM\_DRAM} as \emph{straw-man PIM buddy allocator} for brevity.}, challenges still remain that must be addressed to develop a fast and scalable PIM memory allocator. We detail the two limitations of the straw-man PIM buddy allocator.

First, the inherent design of the buddy allocator causes latency to increase when the heap size expands or the minimum (de)allocation size decreases. The buddy allocator manages the heap using a tree data structure, where larger heap sizes or smaller (de)allocation sizes increase the tree’s depth, requiring deeper tree traversals. This results in longer metadata search times and higher latency during memory (de)allocations. In UPMEM-PIM’s existing scratchpad-based buddy allocator (\texttt{buddy\_alloc()}), the maximum heap size in the scratchpad (WRAM) is capped at 64 KB, with the minimum (de)allocation size set to 32 bytes, resulting in a maximum of a 10-level buddy tree traversal ($log_2(32 \ KB \ / \ 32 \ B) = 10$). Consequently, the baseline scratchpad-based buddy allocator's memory allocation latency is low enough to avoid becoming a performance bottleneck. With our straw-man PIM buddy allocator, however, the heap size can be orders of magnitude larger than the PIM core's scratchpad space, leading to a maximum 20-level tree traversal with heap size of 32 MB and causing significant increases in memory allocation latency. \fig{fig:fast_and_slow_paths} compares the average memory allocation latency of our straw-man PIM buddy allocator over various heap sizes and minimum memory (de)allocation sizes. Compared to a 2 KB allocation in a 32 KB heap, a 32 B allocation over a 32 MB heap experiences up to a 12$\times$ slowdown, highlighting the scalability challenges of our straw-man PIM buddy allocator.

\begin{figure}[t!] \centering
  \includegraphics[width=0.475\textwidth]{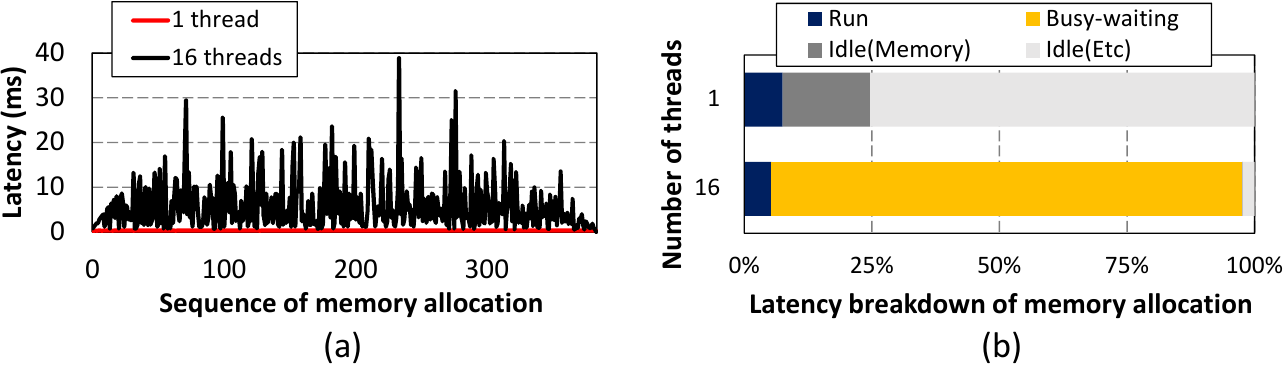} 
  \caption{(a) 
  Changes in memory allocation latency when an UPMEM-PIM program instantiates $N$ threads and each thread requests consecutive memory allocations using the straw-man PIM buddy allocator. (b) Breakdown of memory allocation latency observed across all memory allocation requests in (a). }
  \label{fig:thread_contention}
  \vspace{-1.3em}
\end{figure}

The second challenge of the straw-man PIM buddy allocator is its susceptibility to thread contention. The UPMEM-PIM programming model employs fine-grained multithreading within each PIM core (DPU) to achieve high performance. When multiple threads simultaneously request (de)allocation within a given timeframe, the buddy allocator -- protected by a mutex -- can only process one thread's allocation and deallocation request at a time, forcing the remaining threads to be stuck in a busy-waiting state. In \fig{fig:thread_contention}, we show the effect of thread contention on memory allocation performance. Under a 16-thread execution scenario, severe thread contention causes significant fluctuations in memory allocation latency, unlike in single-thread memory allocation where the latency remains fairly stable as shown in \fig{fig:thread_contention}(a). This behavior occurs because it is busy-waiting for its turn to acquire the mutex and have its allocation request serviced (\fig{fig:thread_contention}(b)).
\section{\pimmalloc: A Fast and Scalable Dynamic Memory Allocator for PIM}
\label{sect:proposed}

We present two different versions of \pimmalloc. \pimmallocSW effectively addresses the limitations of the straw-man PIM buddy allocator purely through software optimizations, allowing it to be readily deployable on existing PIM devices. To achieve further performance gains, we propose \pimmallocHW, a hardware/software co-design incorporating lightweight modifications to the PIM core.

\subsection{Software-only \pimmallocSW}
\label{sect:proposed_sw_only}
\begin{figure}[t!] \centering
%\vspace{-1.3em}
  \includegraphics[width=0.475\textwidth]{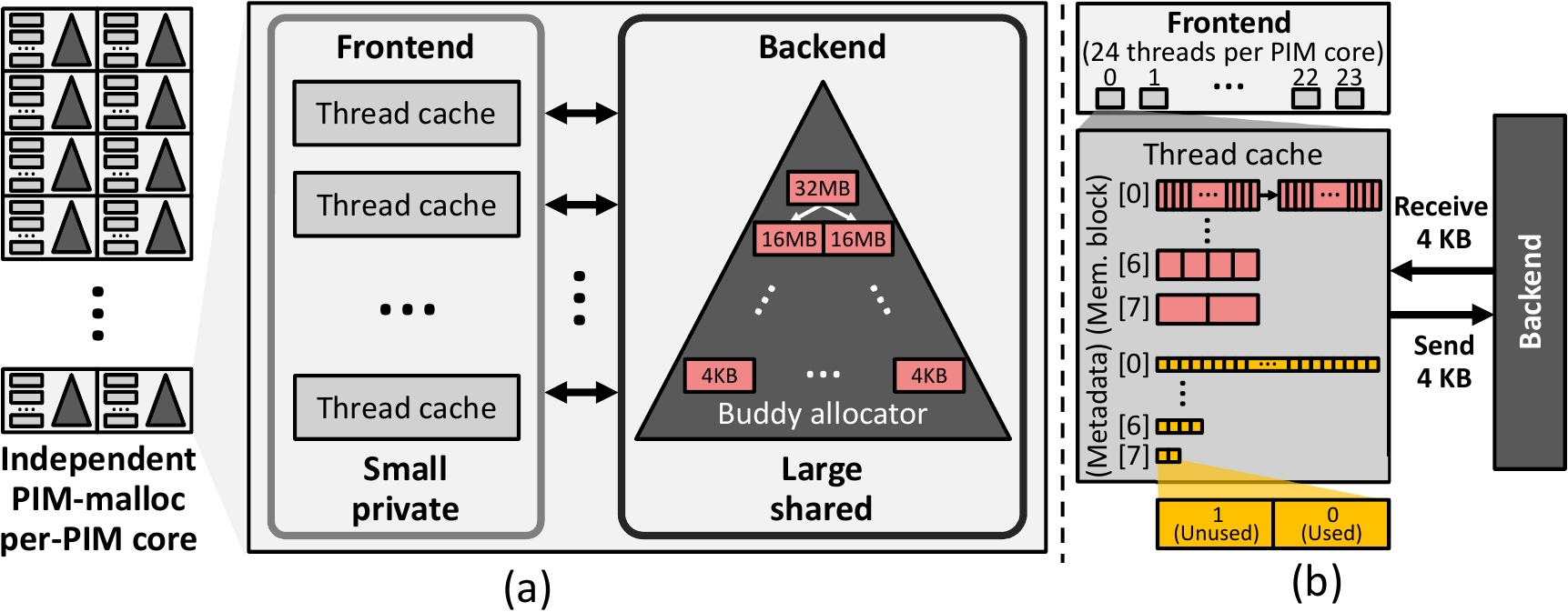} 
  \caption{(a) High-level overview of our \pimmallocSW design and (b) its implementation details.}
  \label{fig:pim_malloc_sw}
  \vspace{-1.3em}
\end{figure}

\textbf{Design principle.} To address limitations in the straw-man PIM buddy allocator, \mbox{\pimmallocSW} employs a hierarchical memory allocation scheme, as illustrated in \mbox{\fig{fig:pim_malloc_sw}}(a): (1) At the frontend, a per-thread \emph{thread cache} manages small memory (de)allocation requests (between 16 bytes and 2 KB). (2) For allocation requests that exceed 2 KB, \mbox{\pimmallocSW} employs the \emph{buddy allocator} in the backend.

\begin{figure*}[t!] \centering
\includegraphics[width=0.90\textwidth]{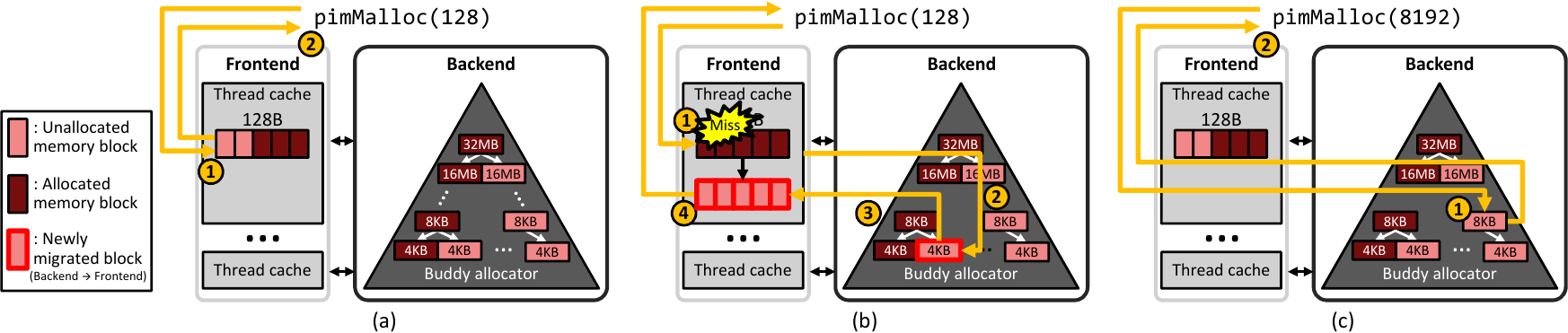}
    \caption{Workflow of \pimmallocSW under several key scenarios: (a) thread cache hit, (b) thread cache miss, and (c) thread cache bypass.}
  \vspace{-1.3em}
	\label{fig:pim_malloc_sw_workflow}
\end{figure*}

In \pimmallocSW, thread caches are memory pools exclusively assigned to each thread. Since each PIM core in UPMEM-PIM can support up to 24 concurrent threads, \pimmallocSW allows for up to 24 thread caches per PIM core (\fig{fig:pim_malloc_sw}(b)).
This design enables each thread to efficiently manage small memory (de)allocations independently, without requiring mutex acquisition. While \pimmallocSW's buddy allocator still uses a shared mutex for multi-threaded allocations and does not support exclusive per-thread memory pools, it benefits from reduced lock contention due to our hierarchical memory allocator design. Specifically, the frontend allocator can filter out allocation requests, reducing the load on the backend allocator. Another key advantage of our hierarchical design is its ability to reduce the buddy allocator’s metadata overhead compared to the single-level buddy allocation algorithm used in the straw-man PIM buddy allocator, which handles both small and large allocation requests. Reducing the metadata tree depth from 20 ($log_2(32 \ MB \ / \ 32 \ B) = 20$) to 13 ($log_2(32 \ MB \ / \ 4 \ KB) = 13$) substantially decreases the buddy allocator's tree traversal overhead. This improvement enhances (de)allocation  speeds for large memory allocations.

Since \pimmallocSW employs a hierarchical allocator design with private thread caches and a shared buddy allocator, one might wonder why we limit our allocator hierarchy to only two layers. Why not adopt a deeper hierarchical allocator design, like TCMalloc~\cite{tcmalloc}, which features four allocator layers and incorporates more sophisticated heap management techniques, such as prefetching memory blocks across different layers? Although this approach might seem advantageous from a performance standpoint, adapting the complex memory allocators to the PIM is impractical for several reasons.

First, PIM devices are fabricated on DRAM process (e.g., $\ge$20 nm DRAM for UPMEM-PIM~\cite{upmem_hotchips}), which imposes severe constraints on their microarchitecture and on-chip storage. For instance, the mere 24 KB of instruction memory (IRAM) in UPMEM-PIM renders complex memory allocators like TCMalloc (approximately 60K C++ lines) infeasible. In contrast, \pimmalloc, with only about 1,000 lines of code, is well-suited for the tight resource budget of PIM devices. Additionally, current PIM systems lack essential system software support like dynamic thread launch, further hindering the implementation of sophisticated memory allocators. Given these challenges, our design aims for a streamlined and efficient approach tailored to the unique resource constraints of PIM devices.

\textbf{Implementation.} \fig{fig:pim_malloc_sw}(b) details the implementation of the thread cache and its interaction with the buddy allocator. The thread cache manages small-sized memory (de)allocations using 4 KB memory blocks provided by the buddy allocator. When free memory blocks are exhausted, the thread cache issues allocation requests to the buddy allocator. Conversely, when all memory sub-blocks within a 4 KB block in the thread cache are freed by the PIM program, they are merged back into a 4 KB block and returned to the buddy allocator.

In our design, each thread cache contains eight linked lists to efficiently manage memory (de)allocations of specific sizes. For instance, the linked list at index 0 handles 16 B (de)allocations, while the linked list at index 7 manages 2 KB (de)allocations. Each linked list can hold multiple 4 KB memory blocks received from the buddy allocator, which are divided into smaller sub-blocks managed by that list (e.g., a single 4 KB block is divided into two 2 KB sub-blocks in the linked list at index 7). 
This sub-block size, termed the \emph{size class}, ensures efficient management and allocation of small memory requests. By maintaining a pool of fixed-size sub-blocks at the private thread caches, \pimmallocSW can quickly fulfill small-sized memory allocation requests with \texttt{O(1)} latency. A key benefit of this size class-based approach is the elimination of external fragmentation within the thread caches. By design, each linked list functions as an independent pool of fixed-size memory chunks. The state of one pool (for instance, a large number of free 32 B chunks) therefore has no bearing on requests for another, such as a 1 KB allocation, because they are serviced by entirely separate pools. This design choice effectively mitigates the fragmentation problem within the thread caches. To track the allocation status of each sub-block in thread caches, we assign dedicated-bit metadata for each sub-block. As shown in \fig{fig:pim_malloc_sw}(b), each of the eight linked lists maintains its own metadata bitmap (yellow-colored box) to track the allocation status of its sub-blocks. For example, to handle a 16 B allocation request, \pimmallocSW searches the metadata bitmap at index 0 for a bit indicating an available sub-block (i.e., a bit with a value of 1). Once it finds such a bit, \pimmallocSW calculates and returns the address of the corresponding 16 B sub-block (red-colored box) based on the bit’s position (offset). Similarly, a 2 KB request is processed using the metadata bitmap at index 7. The buddy allocator is implemented identically to the straw-man PIM buddy allocator described in \sect{sect:dse_buddy_alloc}, featuring a software-managed metadata buffer to cache recently accessed buddy allocator metadata -- except that the buddy allocator's tree depth is reduced from 20 to 13.

\begin{table}[t!]
\vspace{0.5em}
  \centering
  \caption{\pimmallocSW API functions for memory management.}
\scriptsize
\vspace{-0.3em}
\resizebox{0.475\textwidth}{!}{%
  \begin{tabular}{c c}
    \hline
        \textbf{API}             & \textbf{Semantics} \\
    \hline
    \hline
   	\texttt{void initAllocator(int heap size, }	& Initializes the heap region\\
				\texttt{int sizeClasses[])}& and the size classes\\
    \hline                              
        	              	&          Returns the address of            \\
        \texttt{void* pimMalloc(int size)}      &  a newly allocated \\
				&  memory block of \texttt{size} bytes \\
    \hline
   	\texttt{void pimFree(int ptr)}	&  Deallocates the memory block \\
                                &  based on the \texttt{ptr}       \\
	\hline
  \end{tabular}%
  }
\vspace{-1.6em}
  \label{tab:api}
\end{table}

\textbf{Software interface.} \pimmallocSW provides UPMEM-PIM compatible APIs that enable PIM programmers to initialize and manage heap memory (\tab{tab:api}). Before running a PIM program, \pimmallocSW requires initialization, which involves several key steps. First, it resets the metadata of the allocator to ensure a clean and consistent state for memory management. Additionally, the initialization process pre-populates the thread caches with free memory blocks (a single 4 KB block for each linked list within the thread cache), allowing immediate memory allocation when requested. This proactive strategy reduces initial allocation latency and improves system responsiveness. Since initialization is a one-time operation, a designated thread (i.e., the thread with the thread ID of `0') handles this process. After initialization, each thread can independently use \texttt{pimMalloc()} and \texttt{pimFree()}.

\textbf{Workflow.} \fig{fig:pim_malloc_sw_workflow} illustrates the workflow of  \pimmallocSW during a \texttt{pimMalloc()} operation. For simplicity, the explanation of the \texttt{pimFree()} operation is omitted, as it follows a similar logic. The workflow of \pimmallocSW can be divided into three key scenarios as follows:

\emph{(Case \#1: thread cache hit)} This scenario represents the fastest \texttt{pimMalloc()} operation within \pimmallocSW. When a 128-byte  request is received, the allocator examines the linked list in the thread cache designated for 128-byte allocations. If a free 128-byte sub-block is available (\circled{1} in  \fig{fig:pim_malloc_sw_workflow}(a)), \pimmallocSW immediately returns this sub-block, fulfilling the allocation request (\circled{2}).

\emph{(Case \#2: thread cache miss)} If the 128-byte linked list contains no free sub-blocks (\circled{1} in \fig{fig:pim_malloc_sw_workflow}(b)), \pimmallocSW requests a 4 KB memory block from the buddy allocator (\circled{2}). This newly allocated 4 KB block is added to the thread cache's linked list and subdivided into 128-byte sub-blocks (\circled{3}). Finally, one of these 128-byte sub-blocks is returned to fulfill the original memory allocation request (\circled{4}).

\emph{(Case \#3: thread cache bypass)}  This scenario involves an allocation request exceeding the thread cache's largest size class, i.e., the sub-block size. For example, when a 8 KB block is requested, the allocator bypasses the thread cache entirely, as its operational range is limited to allocations between 16 bytes and 2 KB. Instead, \pimmallocSW directly queries the buddy allocator for the 8 KB allocation (\circled{1} in \fig{fig:pim_malloc_sw_workflow}(c)). The result from the buddy allocator is returned to the user program to fulfill the memory allocation request (\circled{2}).

%\vspace{1.0em}
\subsection{Hardware/Software Co-Designed \pimmallocHW}
\label{sect:proposed_hw_sw}

\begin{figure}[t!] \centering
%\vspace{-1.3em}
  \includegraphics[width=0.475\textwidth]{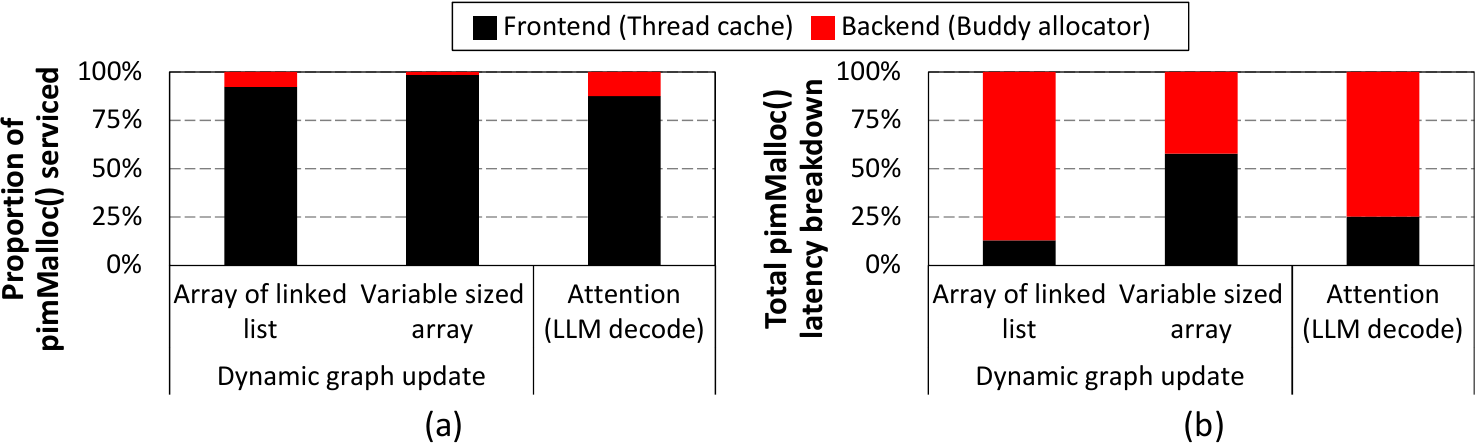} 
  \caption{Characterization of \mbox{\pimmallocSW}'s memory allocation during dynamic graph update and an LLM's attention layer computation: (a) the proportion of \texttt{pimMalloc()} requests serviced at each level and (b) a breakdown of aggregate \texttt{pimMalloc()} latency, based on where the allocation request was serviced.
  }
  \label{fig:level_0_1_breakdown}
  \vspace{-1.3em}
\end{figure}

\textbf{Limitations of \pimmallocSW.} While \pimmallocSW was able to achieve a remarkable speedup over the straw-man PIM buddy allocator, the unique hardware/software constraints of the PIM system pose significant challenges for further improving \pimmallocSW via software-only optimizations. As an example, consider \pimmallocSW's software-managed metadata buffer which is used to cache recently accessed metadata and exploit locality. For the simplicity of its design, the baseline implementation of \pimmallocSW employs a coarse-grained metadata buffer management strategy where a miss at the software-managed buffer leads to flushing this buffer and a subsequent reloading of the required metadata from DRAM. To address this inefficiency, we also explored a fine-grained metadata buffer management mechanism based on an LRU replacement policy, implemented entirely in software. However, this approach encountered several challenges stemming from the inherent computational limitations of PIM. Specifically, such fine-grained buffer management introduces significant software overhead, negating the benefits of reduced DRAM transfers via intelligent caching decisions. As such, our analysis reveals that, while the aforementioned LRU-based metadata buffer management effectively reduces metadata transfers from DRAM, the associated computational overhead results in negligible or even diminished performance gains (fine‑grained buffer management shows a 29\% performance degradation in the microbenchmark described in \sect{sect:methodology}, where 16 threads request 4 KB allocations), which led to our baseline, coarse-grained metadata buffer management in \pimmallocSW. Overall, the difficulty of further improving \pimmallocSW's performance purely via software-level optimizations motivates our hardware/software co-design approach, aka \pimmallocHW, detailed below.

{\bf Prior work on HW-accelerated memory allocators and its differences vs. \pimmallocHW.} Several studies have explored hardware-based approaches to optimize dynamic memory allocators in CPUs~\cite{maas2018hardware, chang1996high, cam1999high, li2006page, von1975simple, mallacc, memento}. Notably, Mallacc~\cite{mallacc} augments the processor core with a \emph{dedicated hardware cache} to enhance the performance of TCMalloc's hierarchical memory allocator design. This work primarily targets the frontend memory allocator (per-CPU cache) of TCMalloc, which is based on the observation that most allocation requests can be satisfied at this layer. One might wonder whether directly applying the approaches suggested in Mallacc~\cite{mallacc} to PIM-malloc, which aims to accelerate \pimmallocSW's frontend memory allocator, could yield performance benefits. However, our characterization of \pimmallocSW revealed unique properties in its two-layer PIM memory allocator design that necessitate a fundamentally different approach than what was taken in Mallacc~\cite{mallacc}. 

\begin{figure}[t!] \centering
%\vspace{-1.3em}
  \includegraphics[width=0.45\textwidth]{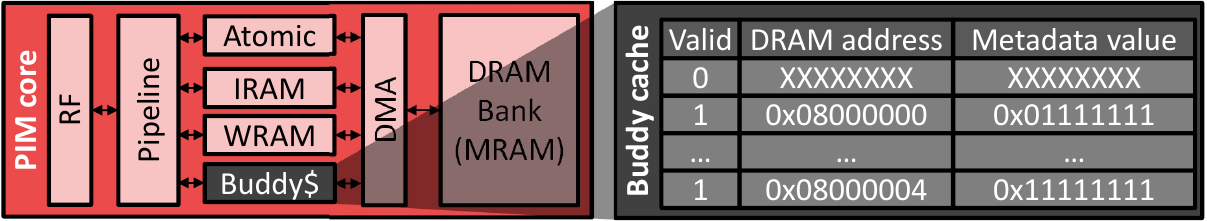}
  \caption{High-level overview of \pimmallocHW incorporating the hardware-based buddy cache microarchitecture.}
  \label{fig:codesign_top_down}
  \vspace{-1.3em}
\end{figure}

In \fig{fig:level_0_1_breakdown}, we analyze memory allocation behavior for our workloads (see \sect{sect:methodology} for its implementation details) and provide several key insights. First, on average, 93\% of allocation requests are satisfied by \pimmallocSW's frontend (\fig{fig:level_0_1_breakdown}(a)). Second, although the backend allocator serves far fewer requests, it accounts for 68\% of the total allocation latency (\fig{fig:level_0_1_breakdown}(b)). This is in stark contrast to TCMalloc, which consumes 53\% of its memory allocation time in the frontend memory allocator~\cite{zhou2024characterizing}, unlike \pimmallocSW's frontend allocator which contributes only 32\% (\fig{fig:level_0_1_breakdown}(b)). Such phenomenon arises from differences between \pimmalloc and TCMalloc. TCMalloc employs a four‑layer hierarchical memory allocator with advanced memory‑pool management techniques, such as memory‑block prefetching, whereas \pimmalloc lacks these features because of PIM constraints, as discussed in \sect{sect:proposed_sw_only}. Thus, we design \pimmallocHW to improve the performance of the backend allocator, unlike the prior approach~\cite{mallacc} which focuses on accelerating the top-level memory allocator.

\textbf{\pimmallocHW overview.} \pimmallocHW alleviates the performance degradation at \pimmallocSW's buddy allocator by employing a hardware-based \emph{buddy cache}, one for each PIM core (\fig{fig:codesign_top_down}). Conceptually, the role of our buddy cache is identical to \pimmallocSW's software-managed metadata buffer: it caches recently accessed buddy allocator metadata for fast tree traversals, but manages the cache in a fine-grained manner based on a hardware-based LRU replacement policy for high performance.

The buddy cache is implemented as a fully-associative cache using Content-Addressable Memory (CAM). Each entry consists of a valid bit (1 bit), the DRAM address of the metadata (4 bytes) serving as the tag, and the metadata value (4 bytes). Our baseline buddy cache utilizes a 16-entry CAM structure (4 B metadata per entry, 16 entries in total) because storing 64 B of metadata per cache was sufficient to capture most access locality (\fig{fig:eval_sensitivity_buddycache_size} details sensitivity to this design parameter). The buddy cache is managed using an LRU replacement policy for efficient metadata retrieval and eviction. \pimmallocHW extends the Instruction Set Architecture (ISA) of our PIM design with the following four instructions to interface the buddy cache with the runtime system:

\begin{itemize}
\item \texttt{init\_bc}: resets and initializes the buddy cache
\item \texttt{lookup\_bc}: conducts a tag lookup over the buddy cache using the DRAM address associated with the stored metadata 
\item \texttt{read\_bc}: read a (metadata) value from the buddy cache based on the input index
\item \texttt{write\_bc}: write a (metadata) value into the buddy cache based on the input index
\end{itemize}

Below we detail how these ISA extensions are utilized to enable high-performance buddy allocator metadata retrieval.  

\begin{figure}[t!] \centering
%\vspace{-1.3em}
  \includegraphics[width=0.475\textwidth]{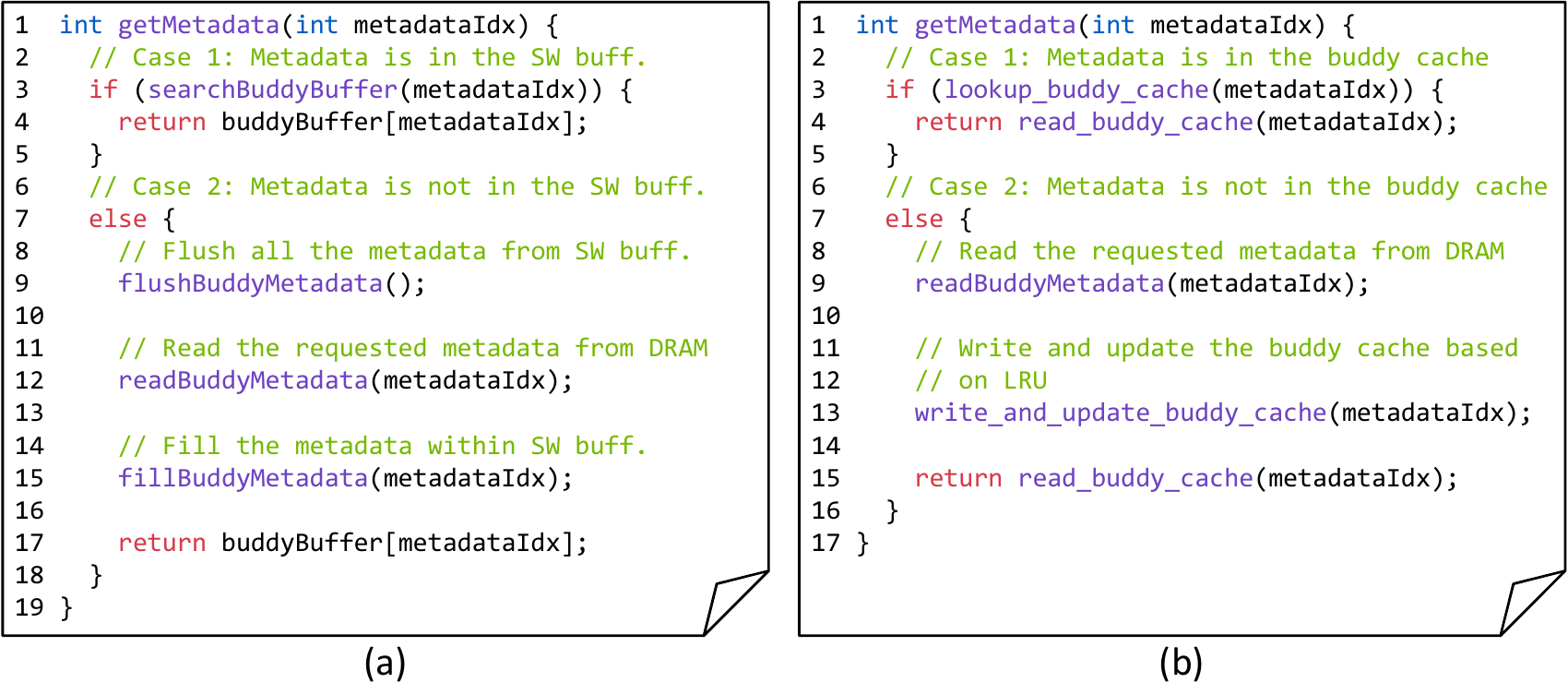} 
  \caption{Code snippets that handle the metadata retrieval process when using  (a) \pimmallocSW and (b) \pimmallocHW.}
  \label{fig:code_sw_versus_hwsw}
 \vspace{-1.3em}
\end{figure}

\textbf{HW/SW interface for the buddy cache.}  \fig{fig:code_sw_versus_hwsw} shows code snippets of the \texttt{getMetadata()} function used as part of implementing the buddy allocator in both \pimmallocSW (\fig{fig:code_sw_versus_hwsw}(a)) and \pimmallocHW (\fig{fig:code_sw_versus_hwsw}(b)). This function retrieves the metadata of a given tree node (\texttt{metadataIdx}) accessed during buddy tree traversal. Initially, both implementations search for the requested metadata in their respective on-chip SRAM storage: (1) \pimmallocSW checks the software-managed metadata buffer stored inside the scratchpad (\texttt{searchBuddyBuffer()}), while (2) \pimmallocHW checks the hardware-based buddy cache (\texttt{lookup\_buddy\_cache()}) by using the \texttt{lookup\_bc} instruction. In the case of a cache hit (case 1, line 2), both implementations return the metadata directly from their respective caches, with \pimmallocHW utilizing the \texttt{read\_bc} instruction as part of the execution of the \texttt{read\_buddy\_cache()} function. However, cache misses (case 2, line 6) can also occur which trigger distinct handling procedures. In \pimmallocSW, the software-managed metadata buffer is flushed (\fig{fig:code_sw_versus_hwsw}(a), line 9) and the required metadata is fetched from DRAM (\fig{fig:code_sw_versus_hwsw}(a), line 12) to populate the metadata buffer with both the requested and adjacent metadata (\fig{fig:code_sw_versus_hwsw}(a), line 15). As for \pimmallocHW, the hardware-based buddy cache facilitates the implementation of a more fine-grained and sophisticated cache replacement policy. Specifically, upon a cache miss, \pimmallocHW fetches only the requested metadata from DRAM (\fig{fig:code_sw_versus_hwsw}(b), line 9) and updates the buddy cache by evicting just the LRU entry using the \texttt{write\_bc} instruction (\fig{fig:code_sw_versus_hwsw}(b), line 13). Compared with \pimmallocSW, this approach not only minimizes metadata related data transfers but also increases the buddy cache hit rate.

\begin{figure}[t!] \centering
%\vspace{-1.3em}
  \includegraphics[width=0.475\textwidth]{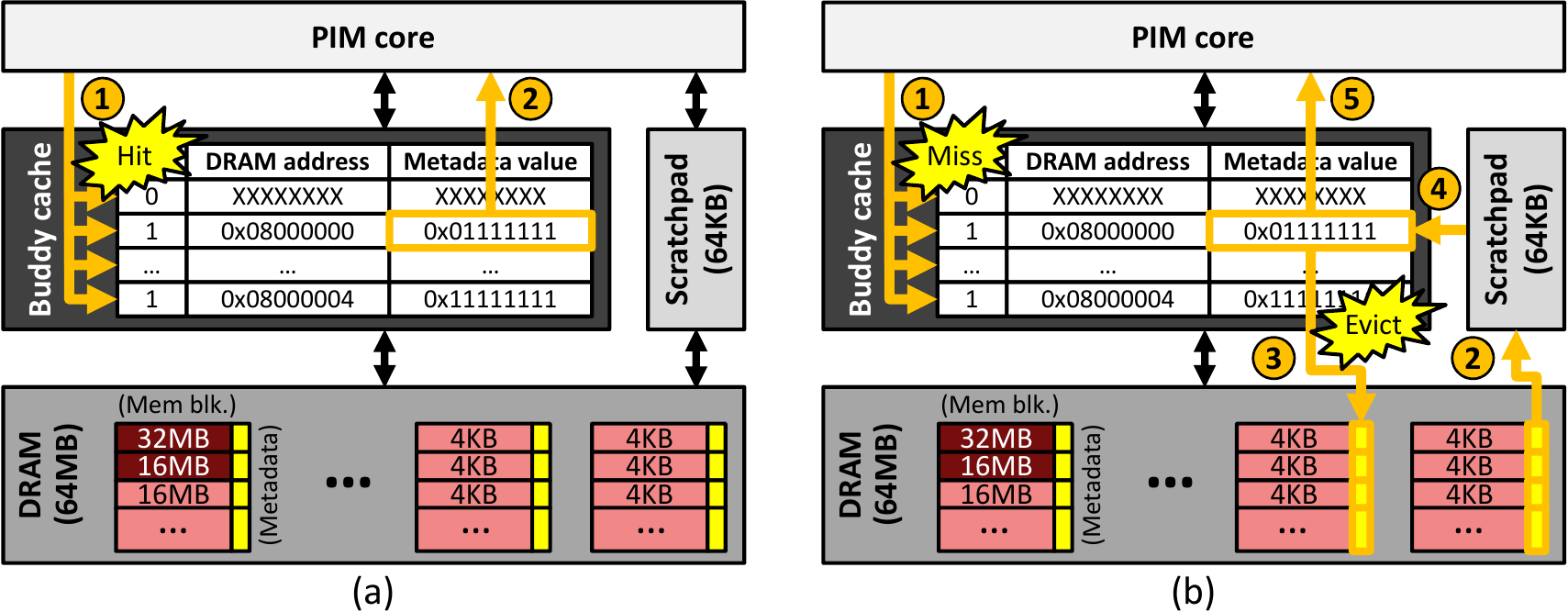} 
  \caption{Workflow of \pimmallocHW during (a) a buddy cache hit (\fig{fig:code_sw_versus_hwsw}(b), line 3-5) and (b) a miss (line 7-16). 
  }
  \label{fig:buddycache_workflow}
  \vspace{-1.3em}
\end{figure}

\textbf{Workflow.} \fig{fig:buddycache_workflow} illustrates the workflow of \pimmallocHW. Initially, \pimmallocHW utilizes the \texttt{lookup\_bc} instruction to check whether the buddy cache stores the metadata associated with the corresponding DRAM address (\circled{1}). If the buddy cache returns a positive value, it corresponds to a buddy cache hit scenario (\fig{fig:buddycache_workflow}(a)), whereas a negative value return corresponds to a cache miss (\fig{fig:buddycache_workflow}(b)). When buddy cache hits, the memory allocator executes a \texttt{read\_bc} instruction to retrieve the corresponding metadata (\circled{2}). Conversely, during a buddy cache miss, the memory allocator first retrieves the metadata from DRAM and temporarily stores it inside the on-chip scratchpad (\circled{2}), evicts the LRU entry from the buddy cache (\circled{3}), and then fills in  the newly fetched metadata into the now vacant buddy cache entry using the \texttt{write\_bc} instruction (\circled{4}). This ensures that the requested metadata is readily available from the buddy cache for subsequent metadata access by the PIM core (\circled{5}).

\section{Methodology}
\label{sect:methodology}

{\bf Characterization using a real PIM system.}
All the experiments conducted in \sect{sect:motivation} and \sect{sect:proposed} (with the exception of \fig{fig:thread_contention}(b), based on simulator~\cite{upimulator}, due to the lack of profiling tools) were conducted on a real UPMEM-PIM system. This system includes an Intel Xeon Gold 5222 CPU as the host processor, equipped with two channels of DDR4-3200 and four channels of DDR-2400-based UPMEM-PIM DIMMs, containing a total of 512 PIM cores.

\textbf{Evaluation using cycle-level simulation.} We use the open-source UPMEM-PIM cycle-level simulator, uPIMulator~\cite{upimulator}, for all our evaluation in \sect{sect:evaluation} that compare the performance of straw-man PIM buddy allocator, \pimmallocSW, and \pimmallocHW. For \pimmallocHW, we faithfully model the behavior of the buddy cache, the support for ISA extensions, and their HW/SW interface inside uPIMulator. We adopt the default simulation configuration provided by uPIMulator~\cite{upimulator} and configure \pimmalloc as follows: a 32 MB heap size; allocation size classes defined by powers of two (with a minimum of 16 B and a maximum of 2 KB, total of 8 size classes); a buddy cache size of 64 B; and a buddy cache access latency of 1 PIM core cycle.

\textbf{Microbenchmark.} 
In \sect{sect:eval_6_1}, we evaluate PIM-malloc's standalone performance by comparing the average latency of the straw-man PIM buddy allocator, \pimmallocSW, and \pimmallocHW. Our microbenchmark performs a series of \texttt{pimMalloc()} calls to measure latency while varying the number of threads, allocation size, etc.

\textbf{Benchmark for PIM-malloc use case.} To demonstrate a practical use case of our \pimmalloc design, we evaluate (1) \emph{dynamic graph update} and (2) \emph{attention layer in LLM inference}. For the dynamic graph update, we follow the evaluation methodology used in prior work~\cite{awad2020dynamic, sagabench, basak2021improving, terrace, faimgraph, aspen}. Due to the limited availability of open-source datasets that model dynamic graph update behaviors in real-world applications, these prior studies perform experiments using synthetic datasets. Specifically, nodes or edges of a static graph dataset are randomly sampled; the sampled portion represents the newly added graph dataset, while the unsampled portion represents the existing graph dataset. As such, we set the size ratio of the newly added dataset to the existing dataset at 1:2 and perform random sampling on the edges. We use the loc-gowalla~\cite{cho2011friendship} dataset, employed in the PrIM~\cite{prim} benchmark suite, as our graph dataset. It is important to note that the performance of dynamic graph updates varies by data structure. Therefore, we evaluate two representative structures from prior work: an array of linked lists~\cite{faimgraph} and a variable-sized array~\cite{hornet}. The former uses a constant allocation size (we assume 256 B) because its edge-storing elements are fixed-size arrays. The latter manages all edges for a node in a single power-of-two-sized array, resulting in allocation sizes that vary from 64 B to 32 KB in our dataset.
For the attention layer, we adopt the Llama‑2 7B~\cite{llama2} model configuration to calculate the KV cache size and extend the PrIM~\cite{prim} GEMV benchmark to execute the attention computation on PIM. To handle the dynamic growth of the KV cache, our PIM kernel is designed to allocate a new 512 B memory block whenever the existing space is exhausted. We evaluate LLM inference performance by generating a trace using simulation results from uPIMulator~\cite{upimulator}. This trace represents a workload of 100 total requests arriving at a rate of 10 requests per second, each with an input prompt of 128 tokens and an output of 256 tokens. This trace is then processed by LLMServingSim~\cite{llmservingsim}, a system-level simulator for LLM inference serving.

%\vspace{-1em}
\section{Evaluation}
\label{sect:evaluation}

\subsection{Microbenchmark based Analysis}
\label{sect:eval_6_1}

\begin{figure}[t!] \centering
%\vspace{-1.3em}
  \includegraphics[width=0.475\textwidth]
  {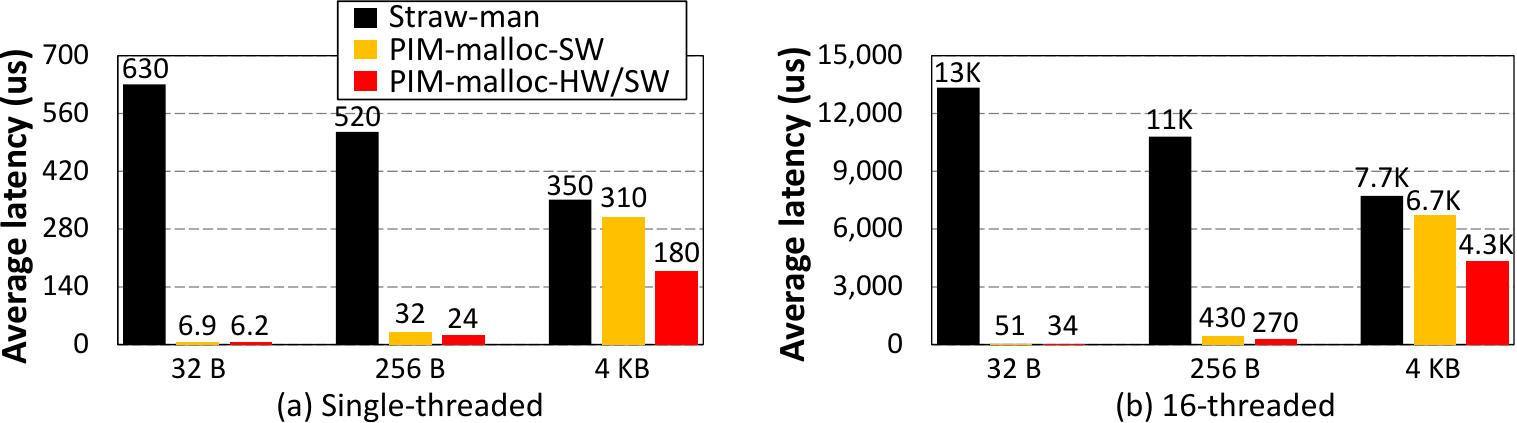} 
    \caption{Average memory allocation latency when executing our microbenchmark employing (a) a single thread (no lock contention) and (b) 16 threads (lock contention) for memory allocation requests.}
  \label{fig:eval_malloc_latency}
  \vspace{-1.0em}
\end{figure}

{\bf Average memory allocation latency.}
We compare memory allocation latency across various allocators using the microbenchmark described in \sect{sect:methodology}. The microbenchmark uses either one (\fig{fig:eval_malloc_latency}(a)) or 16 threads (\fig{fig:eval_malloc_latency}(b)), each making 128 allocation requests of a fixed size. The straw-man allocator has the highest latency, since it needs to traverse a large tree structure for every request. For smaller allocations (32 B and 256 B), both \pimmalloc designs rely on a thread cache to achieve \texttt{O(1)} latency, yielding significant speedups. \pimmallocHW provides further speedup over \pimmallocSW for 32 B and 256 B allocations because, in cases where the thread cache is empty, its hardware-based buddy cache accelerates tree traversal. For allocation requests of 4 KB, these are serviced directly by the buddy allocator for all three design points, so their performance differences primarily come from how efficiently each design's buddy allocator is implemented. \pimmallocHW{}’s hardware-based cache reduces 4 KB allocation latency by 39\% compared to \pimmallocSW. Overall, \pimmallocSW achieves a 66$\times$ speedup over the straw-man allocator, while \pimmallocHW outperforms \pimmallocSW by 31\%.

\setlength{\columnsep}{8pt}
\setlength{\intextsep}{0pt}
\begin{wrapfigure}{r}{0.475\linewidth}
%\vspace{0.2em}
\includegraphics[width=0.90\linewidth]{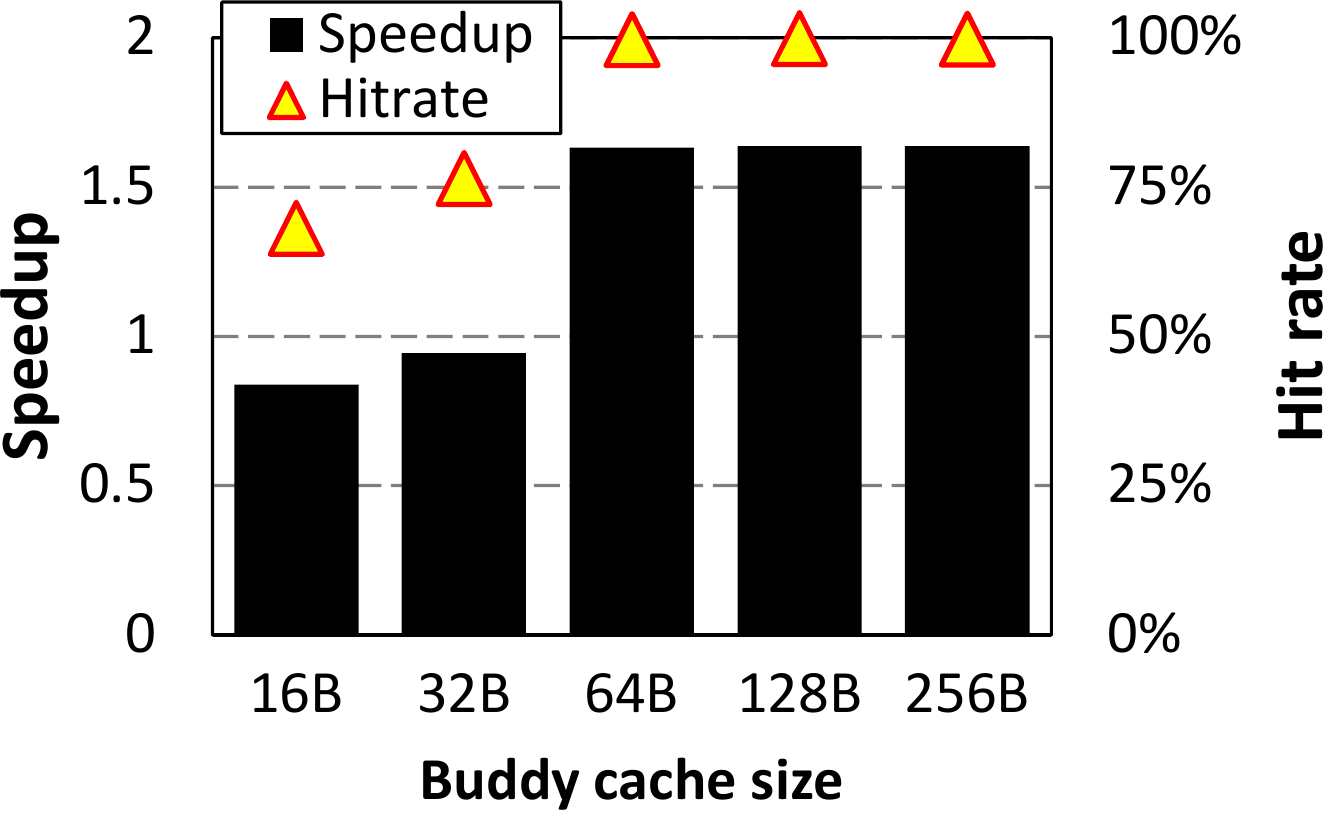}
\captionof{figure}{\pimmallocHW's speedup over \pimmallocSW and the corresponding buddy cache hit rate when the buddy cache size is changed. Results are collected using the same microbenchmark utilized in \fig{fig:eval_malloc_latency} with 16 threads and 4 KB  request per \texttt{pimMalloc()}.}
\label{fig:eval_sensitivity_buddycache_size}
\end{wrapfigure}
%\par

\textbf{\pimmallocHW's sensitivity to buddy cache size.}
\fig{fig:eval_sensitivity_buddycache_size} shows \pimmallocHW's speedup over \pimmallocSW and the corresponding buddy cache hit rate. Both the speedup and the buddy cache hit rate saturate beyond a cache size of 64 B, which can be explained by the metadata size of the buddy cache. In our buddy allocator implementation, each tree node uses 2 bits of metadata to track the allocation status of a node (tracking three states, i.e., fully allocated, partially allocated, unallocated), so a 64 B buddy cache can store metadata for up to 256 node elements ($\frac{64 \ bytes \ \times \ 8 \ bits/byte}{2 \ bits/element} = 256 \ elements$). 64 B capacity was shown to be sufficient to capture the locality in frequently traversed tree paths, resulting in \pimmallocHW's speedup saturating beyond this design point.

\begin{figure*}[t!] \centering
  \includegraphics[width=0.99\textwidth]{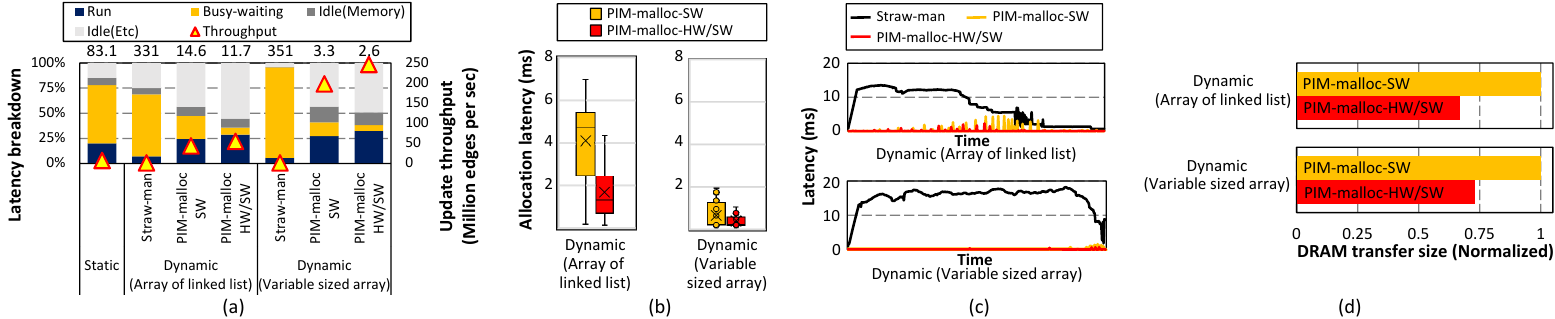} 
  \caption{Experiments of dynamic graph update workload: (a) Graph update throughput (right axis), absolute latency (top), and latency breakdown (left axis), (b) distribution of total \texttt{pimMalloc()} execution time per thread, (c) changes in memory allocation latency during the execution of graph updates, and (d) comparison of DRAM transfer sizes for \pimmallocSW vs. \pimmallocHW.
  }
  \label{fig:eval_graph_update}
  \vspace{-1.3em}
\end{figure*}

\textbf{Comparison of metadata management in \pimmallocSW and \pimmallocHW.}
The buddy allocator traverses a buddy tree to access metadata, creating a non-sequential access pattern as it probes physically non-adjacent entries. This characteristic renders the coarse-grained caching scheme of \pimmallocSW inefficient. This scheme fetches large, often unrelated metadata blocks, which wastes MRAM-to-WRAM bandwidth. Consequently, it suffers from a low buffer hit rate and a high volume of data transfers caused by frequent block replacements. In contrast, the fine-grained approach of \pimmallocHW effectively leverages the buddy allocator's temporal locality through a hardware-based LRU policy. It minimizes unnecessary data transfers by retaining the metadata of frequently revisited parent nodes from upper tree levels in its buddy cache. The results from a 4 KB allocation microbenchmark clearly illustrate this difference. \pimmallocHW achieved a 99\% hit rate, transferring on average only 2-bytes of metadata per \texttt{pimMalloc()} request, whereas \pimmallocSW's 73\% hit rate resulted in 2 KB of data transfer per \texttt{pimMalloc()} request. However, this performance gap narrows under certain conditions, such as with large allocation sizes (e.g., 256 KB) and a low number of requests. This is because the shallow search depth required for large allocations increases metadata reuse within a single coarse-grained block, and the low request frequency makes buffer replacements infrequent.

\subsection{Case Study \#1: Dynamic Graph Updates}
\label{sect:eval_case_study_1}
In this section, we evaluate the performance of dynamic graph updates. \fig{fig:eval_graph_update}(a) compares the performance of implementations based on static (CSR) and dynamic data structures (array of linked lists and variable-sized arrays), where the latter is implemented using different types of dynamic memory allocators. The dynamic data structure, built with the straw-man PIM buddy allocator, exhibits lower performance compared to the static baseline, primarily due to excessive ``busy-waiting''. \pimmallocSW alleviates this bottleneck by employing a lock-free thread cache for small allocations, thereby reducing contention and increasing the computation time ratio (``Run''). \pimmallocHW further improves performance by using a buddy cache that accelerates metadata retrieval, allowing it to outperform the straw-man PIM buddy allocator design. Consequently, \pimmallocHW demonstrates a 7.1$\times$ and 32$\times$ speedup over the static baseline for the array of linked lists and the variable-sized array implementations, respectively. \fig{fig:eval_graph_update}(b) presents the allocation latency distribution during dynamic graph updates. The straw-man allocator suffers from extremely high average latencies, on the order of hundreds of milliseconds, which are omitted from the figure for readability. By comparison, both \pimmallocSW and \pimmallocHW drastically reduce this latency, underscoring that minimizing allocation overhead is critical for achieving high workload performance. \fig{fig:eval_graph_update}(c) shows the changes in memory allocation latency over time during dynamic graph updates. The straw-man PIM buddy allocator consistently exhibits the highest latency, with significant fluctuations caused by frequent inter-thread lock contention. In contrast, both \pimmallocSW and \pimmallocHW maintain relatively stable and lower latencies, while \pimmallocHW consistently outperforms \pimmallocSW. Occasional latency spikes occur in both designs which correspond to thread cache misses, where requests fall back to the slower buddy allocator. These results highlight the significant latency disparity between the frontend and the backend memory allocator. \fig{fig:eval_graph_update}(d) further shows the advantages of \pimmallocHW{}’s fine-grained buddy cache management and LRU replacement policy which reduce aggregate DRAM transfer size by 30\% compared to \pimmallocSW{}’s software-managed metadata buffer.

\begin{figure}[t!] \centering
%\vspace{-1.3em}
  \includegraphics[width=0.35\textwidth]{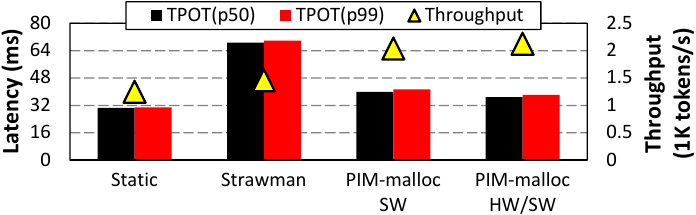} 
  \caption{Comparison of measured throughput and TPOT across four different allocation schemes.}
  \label{fig:eval_attention}
  %\vspace{-1.3em}
\end{figure}

\subsection{Case Study \#2: Attention Layer in LLM Inference}
\label{sect:eval_case_study_2}

Two key metrics measure the performance of LLM serving: (1) \emph{Throughput}, the number of tokens processed per second, and (2) \emph{Time-Per-Output-Token} (TPOT), the latency between generating consecutive tokens. As explained in \sect{sect:why_malloc_for_pim}, managing the KV cache for LLM inference with dynamic allocation enables more efficient memory usage. This allows for more concurrent requests, resulting in higher throughput compared to static allocation (\fig{fig:eval_attention}). Furthermore, reducing the dynamic allocator's latency frees up PIM's computational resources for the attention operation, further improving throughput. Among the dynamic methods, \pimmallocHW delivers the highest throughput (1.7$\times$ improvement over static allocation). However, this approach presents a trade-off: dynamic memory allocation introduces overhead that increases the attention layer's latency, which in turn increases TPOT. This trend is observed consistently across all TPOT percentiles. The straw-man method, for instance, exhibits the highest TPOT among the four schemes due to its high allocation latency. In contrast, \pimmallocSW effectively achieves a lower TPOT than the straw-man approach via its thread cache (frontend) design. Building on this, \pimmallocHW leverages its buddy cache to improve both the 95th and 99th percentile TPOT by another 8\% over \pimmallocSW, resulting in TPOTs that are 21\% and 24\% higher than the static allocation baseline, respectively.

\subsection{Fragmentation Analysis}
\label{sect:fragmentation}
\pimmalloc mitigates fragmentation by employing dedicated memory pools for each size class. However, to reduce initial allocation latency, the current implementation includes an optimization that pre-allocates all thread caches with free 4 KB memory blocks during the \texttt{initAllocator()} call, as discussed in \sect{sect:proposed_sw_only}. While this pre-population strategy can enhance performance, it may also lead to unnecessary memory fragmentation, depending on the workload's allocation patterns. For example, if threads on a PIM core only allocate memory blocks of a specific size, the pre-allocated memory in other size classes remains entirely unused.

\tab{tab:fragmentation} quantifies memory fragmentation ($A$/$U$), measured as the ratio between ``total allocated memory by allocator ($A$)'' and ``total memory requested by program ($U$)'', as defined in~\cite{hoard}, e.g., a fragmentation value higher than 1.0 means that memory is reserved but not effectively used by the program. The dynamic graph update (array of linked lists) workload, which uses only a single size class, exhibits a high fragmentation of 1.95 due to its unused pre-allocated memory blocks. In contrast, fragmentation decreases to 1.72 for the dynamic graph update (variable sized array), which utilizes a variety of block sizes. Interestingly, the LLM attention workload shows even lower fragmentation at 1.66. Although it also uses a single allocation size, its total allocated volume is large enough to amortize the unused pre-allocated blocks. To address this problem, we evaluate a version of our allocator with pre-allocation disabled (\emph{PIM-malloc-lazy} in \tab{tab:fragmentation}). This ``lazy" approach, which allocates memory on-demand rather than pre-filling thread caches, shows a significant reduction in fragmentation across all workloads. These findings highlight a clear trade-off between initial allocation performance and overall memory efficiency. We leave the optimization of this balance, potentially through compiler-based analysis or a user-configurable interface, as future work.

\begin{table}[t!]
%\vspace{0.5em}
  \centering
  \caption{Comparison of memory fragmentation between \pimmalloc (as-is) and \pimmalloc-lazy.}
\scriptsize
\vspace{-0.3em}
\resizebox{0.475\textwidth}{!}{%
  \begin{tabular}{c c c}
    \hline
        \textbf{Workload}             & \textbf{\pimmalloc (as-is)} & \textbf{\pimmalloc-lazy}\\ 
    \hline
    \hline
        Dynamic graph update          & \multirow{2}{*}{1.95}       & \multirow{2}{*}{1.21} \\
        (Array of linked list)        &                             &                       \\
    \hline
        Dynamic graph update          & \multirow{2}{*}{1.72}       & \multirow{2}{*}{1.49} \\
        (Variable sized array)        &                             &                       \\
    \hline
        LLM attention                 & 1.66                        & 1                     \\
    \hline
  \end{tabular}%
  }
\vspace{-1.6em}
  \label{tab:fragmentation}
\end{table}

\subsection{Metadata Storage Overhead of \pimmalloc}
\label{sect:metadata_storage_overhead}
According to our analysis, the metadata storage overhead of \pimmalloc is minimal. Its hierarchical structure reduces the buddy tree depth to 13 compared to a straw-man design, shrinking the metadata to just 4 KB per DRAM bank. \pimmalloc also uses a bitmap for the thread cache's metadata, making its storage cost negligible. Across our workloads, the maximum overhead is 5.1 KB for dynamic graph update (array of linked lists), 5 KB for dynamic graph update (variable-sized array), and 5.2 KB for LLM attention per request.

\subsection{Implementation Overhead}
We use CACTI 7.0~\cite{cacti} with a 32 nm logic process technology to evaluate the area, power, and timing overheads of the buddy cache. The DRAM process used to implement PIM cores is approximately $10\times$ less dense and $3\times$ slower than the logic process~\cite{upmem_hotchips}, so we scale the overhead of our buddy cache accordingly. Our evaluation indicates that the buddy cache adds only 0.019~mm$^2$, a power overhead of 5~mW, and an access latency of less than one PIM core logic cycle.
\section{Discussion}
\label{sect:discussion}

\textbf{Applicability of PIM-malloc to future PIM architectures.} 
We consider two potential architectural evolution paths inspired by recent work by Hyun et al.~\cite{upimulator}, which discussed future PIM capable of exploiting higher parallelism and locality. First, consider a future PIM device with enhanced processing capabilities enabled by DRAM process optimizations. Although such advancements would reduce the absolute latency of \texttt{pimMalloc()}, they would also proportionally accelerate the overall workload, implying that the relative overhead of memory allocation would likely remain a key performance bottleneck. Second, in cache-enabled PIM architectures, even with a general-purpose data cache, a specialized metadata cache, such as our proposed buddy cache, remains necessary. This need stems from a fundamental granularity mismatch: general-purpose caches operate on coarse-grained cache lines (e.g., 64 bytes), which is inefficient for managing the fine-grained metadata used by a buddy allocator. Our buddy cache addresses this by handling metadata at a much finer granularity (e.g., 8 bytes), complementing the general-purpose cache and addressing the memory allocation bottleneck.
%\vspace{-1em}
\section{Related Work}
\label{sect:related}

\textbf{System software support for PIM systems.} Several recent studies have explored system software support for PIM architectures~\cite{affinity_alloc, puma, pim_mmu, pimmmu_cal, simple_pim, pid_comm, transpimlib, transpimlib_arxiv, simdram, umpim, virtualpim, pygim, syncron, mimdram}. SimplePIM~\cite{simple_pim} presents a software framework that enhances UPMEM-PIM system programmability by introducing abstractions akin to those found in distributed frameworks. Syncron~\cite{syncron} provides a hardware/software co-design to support synchronization for PIM. PUMA~\cite{puma} introduces a subarray-level PIM~\cite{seshadri2015fast, seshadri2016buddy, rowclone, simdram, ambit, lisa, pluto, yuksel2024functionally, computedram} aware memory allocator, which supports the memory object to be placed within the specific DRAM subarrays. To the best of our knowledge, this work is the first to systematically address the unique challenges of dynamic memory allocation in  general-purpose PIM systems.

\textbf{Dynamic memory management for parallel computing.} Numerous studies have been conducted to improve the performance of parallel memory (de)allocation across different threads in multi-core CPUs~\cite{jemalloc, hoard, tcmalloc, magazine_allocator} and GPUs~\cite{winter2021dynamic, pham2022dynamic, mccoy2024gallatin, gelado2019throughput, xmalloc}. Hoard~\cite{hoard} proposes a scalable allocator aimed at efficiently handling multi-threaded CPU workloads by introducing a per-thread heap (which each thread can uniquely access). XMalloc~\cite{xmalloc} presents a scalable, lock-free dynamic memory allocator designed for GPUs, addressing the scalability limitations of traditional Compare-And-Swap based allocators by leveraging SIMD-aware allocation strategies and hierarchical caching mechanisms. While prior work shares certain design principles with \pimmallocSW, \pimmalloc uniquely co-designs hardware/software for heterogeneous (host processor and PIM) systems, exploring optimal metadata placement and a lightweight accelerator which distinguish itself  from existing approaches.

\section{Conclusion}
\label{sect:conclusion}
This work addresses the challenges of enabling dynamic memory allocation in general-purpose PIM through a design space exploration of metadata placement and management. Building on these insights, we propose two high-performance allocators tailored for commodity PIM. Our software-only \pimmallocSW outperforms a straw-man PIM buddy allocator by $66\times$, and our hardware/software co-design \pimmallocHW yields an additional $31\%$ performance gain. These results underscore the substantial benefits of optimizing dynamic allocation in PIM systems, paving the way for broader adoption of this emerging technology.

\section*{Acknowledgment}
This work was partly supported by Institute of Information \& Communications Technology Planning \& Evaluation (IITP) grant funded by the Korea government (MSIT) (No.RS-2024-00438851, (SW Starlab) High-performance Privacy-preserving Machine Learning System and System Software), (No.RS-2024-00402898, Simulation-based High-speed/High-Accuracy Data Center Workload/System Analysis Platform), (No.RS-2025-02214652, Development of SoC Technology for AI Semiconductor-Converged Pooled Storage/Memory), (No. 2022-0-01037, Development of High Performance Processing-in-Memory Technology based on DRAM), and IITP under the Graduate School of Artificial Intelligence Semiconductor (IITP-2026-RS-2023-00256472) grant funded by MSIT. The EDA tool was supported by the IC Design Education Center (IDEC), Korea. We also appreciate the support from Samsung Advanced Institute of Technology and Samsung Electronics Co., Ltd (MEM240728\_0001). Minsoo Rhu is the corresponding author.

\bibliographystyle{IEEEtranS}
\bibliography{refs}

\end{document}